\documentclass[12pt,reqno]{amsart}
\usepackage{graphicx}
\usepackage{color}
\usepackage{amsmath,amsfonts,amssymb,amscd,amsthm,amsbsy,epsf}
\usepackage{cite}
\numberwithin{equation}{section}
\newtheorem{thm}{Theorem}[section]
\newtheorem{lem}[thm]{Lemma}
\newtheorem{definition}[thm]{Definition}
\newtheorem{corollary}{Corollary}
\newtheorem*{cor}{Corollary}
\newtheorem{prop}[thm]{Proposition}
\theoremstyle{definition}
\newtheorem*{remark}{Remark}
\def\step#1{\medskip\noindent{\bf #1.}}

\def\D{{\mathcal D}}
\def\F{{\mathcal F}}
\def\L{{\mathcal L}}
\def\real{{\mathbb R}}
\def\wsuprao{{_{1}}}
\def\wsuprat{{_{2}}}
\def\ep{\varepsilon}

\def\tp{\mathrm{p}}
\def\Norm#1{|\!|\!|#1|\!|\!|}
\def\mustar{{\mu_*}}
\begin{document}
%{\bf version June 10, 06} -- 14maxmol.tex

\

\

 \title[On the self-similar asymptotics for non-linear
Maxwell models]
{On the self-similar asymptotics for
 generalized non-linear kinetic
Maxwell models}
 \author{A.V. Bobylev($^*$), C. Cercignani($^{**}$), I.M. Gamba($^{***}$)}

\address{\(^*\) Department of  Mathematics, Karlstad University,
Karlstad, SE-651\,88 Sweden.}
\email{alexander.bobylev@kau.se}

\address{\(^{**}\) Politecnico di Milano, Milano, Italy.}
\email{carcer@mate.polimi.it}

\address{\(^{***}\) Department of Mathematics,
The University of Texas at Austin,
Austin, TX 78712-1082 U.S.A. }
\email{gamba@math.utexas.edu}

\begin{abstract}
Maxwell models for nonlinear kinetic equations have many
applications in physics, dynamics of granular gases, economy, etc.
In the present manuscript we consider such models  from a very
general point of view, including  those with arbitrary polynomial
non-linearities and in any dimension space. It is shown that the
whole class of generalized Maxwell models satisfies properties which
one of them can be interpreted as an operator generalization of
usual Lipschitz conditions. This property allows to describe in
detail a behavior of solutions to the corresponding initial value
problem. In particular, we prove in the most general case an
existence of self similar solutions and study the convergence,  in
the sense of probability measures,  of dynamically scaled solutions
to the Cauchy problem to those self-similar solutions, as time goes
to infinity. The properties of these self-similar solutions, leading
to non classical equilibrium stable states, are studied in detail.
We apply the results to three different specific problems related to
the Boltzmann equation (with elastic and inelastic interactions) and
show that all physically relevant properties of solutions follow
directly from the general theory developed in this paper.
\end{abstract}

 \maketitle

 \tableofcontents

 \baselineskip=18pt
 \section{Introduction}

The classical (elastic) Boltzmann equation with the Maxwell-type
interactions is well-studied in literature
(see \cite{B88,Ce} and references therein). Roughly speaking, this is a
mathematical model of a rarefied gas with binary collisions such that the
collision frequency is independent of the velocities of colliding particles.

Maxwell models of granular gases were introduced relatively recently in
\cite{BCG}
(see also \cite{B-N-K} for the one dimensional case).
Soon after, these models became very
popular among people studying   granular gases (see, for example,
 the book \cite{PB} and
references therein). There are two obvious reasons for this
fact. First,  the
inelastic Maxwell-Boltzmann equation can be essentially simplified by the
Fourier transform similarly to the elastic one \cite{Bo76,BCG} and second,
 solutions to the spatially homogeneous inelastic Maxwell-Boltzmann
equation have a non-trivial self-similar asymptotics, and, in
addition,
 the corresponding self-similar solution has a power-like tail for large
velocities. The latter property was conjectured in \cite{E-B}
 and later proved in
\cite{BC-03,BCT-03} (see also \cite{BCT-06}).
It is remarkable that such an asymptotics is absent in
the elastic case (roughly speaking, the elastic Boltzmann equation has too
many conservation laws). On the other hand,  the self-similar asymptotics
was proved in the elastic case for initial data with infinite energy
\cite{BC-02} by
using other mathematical tools compared to \cite{BC-03}.
 The recently published
exact self-similar solutions \cite{BG-06} for elastic Maxwell
mixtures (also with power-like tails) definitely suggest the
self-similar asymptotics for such {\sl elastic} systems. Finally we
mention recent publications \cite{BN05,To-05, PT05} , where one
dimensional  Maxwell-type models were introduced for applications to
economy models  and again the self-similar asymptotics and
power-like tail were found.

Thus all the above discussed models describe qualitatively different
processes in physics or economy,
however their solutions have a lot in common from the mathematical point of
view. It is also clear that some further generalizations are possible: one
can, for example, include in the model multiple (not just binary)
interactions still assuming the constant (Maxwell-type) rate of
interactions. Will the multi-linear models have similar properties? The
answer to this question is affirmative, as we shall see below.
It becomes clear that there must be
some general mathematical properties of Maxwell models, which, in turn, can
explain properties of  any particular model. Essentially, there must be
just one {\sl main theorem}, from which one can deduce all the
above discussed facts and their possible generalizations. The goal of this
paper is to consider Maxwell models from a  very general point of view and to
establish their key properties that lead to the self-similar asymptotics.

The paper is organized as follows. We introduce in Section 2 three specific
Maxwell models of the Boltzmann equation: {\bf (A)} classical (elastic)
Boltzmann
equation; {\bf (B)} the model  {\bf (A)} in the presence of a thermostat;
 {\bf (C) } inelastic
Boltzmann equation. Then, in Section 3, we perform the Fourier transform and
introduce an equation that includes all the three models as particular
cases. A further generalization is done in Section 4, where the concept of
generalized multi-linear Maxwell model (in the Fourier space) is
introduced. Such models and their generalizations are studied in detail in
Sections 5-10.  The concept of an {\sl $L$-Lipschitz} nonlinear
operator, one of the most important for our approach,
 is explained in Section 4 (Definition~\ref{def4.1}). It is proved
(Theorem~\ref{thm-lip}) that all multi-linear Maxwell models satisfy
the $L$-Lipschitz
condition. This property of the models constitutes a basis for the general
theory.

 The existence and uniqueness of solutions to the initial value
problem is proved in Section 5 (Theorem 5.2). Then we study in Section 6 the
large time asymptotics under very general conditions that are fulfilled, in
particular, for all our models. It is shown that the
$L$-Lipschitz condition leads
 to self-similar asymptotics provided the corresponding
self-similar solution does exist. The existence and uniqueness of
self-similar solutions is proved in Section 7
(Theorem~\ref{thm7.1}). This result can be considered, to some extent, as
 the {\sl main
theorem}  for general Maxwell-type models. Then, in Section 8, we go back to
the multi-linear models of Section 4 and study more specific properties of
their
self-similar solution. We explain in Section 9 how to use our theory for
applications to any specific model: it is shown that the results can be
expressed in terms of just one function $\mu (p), p>0,$
that depends on the  spectral properties of the specific model.

General properties (positivity, power-like tails,  weak convergence
of probability measures, etc.) of the self-similar solutions are
studied in Section 10. This study also includes the case of one
dimensional models, where the Laplace (instead of Fourier) transform
is used.

Finally, in Section 11, we establish in the unified statement
(Theorem 11.1) the main properties of Maxwell models {\bf (A),(B)}
and {\bf  (C)} of the Boltzmann equation. This result is, in
particular, an essential improvement of earlier results of [7] for
the model {\bf (A) } and quite new for the model {\bf (B).}
Applications to one dimensional models are also briefly discussed at
the end of Section 11.

\vskip.5in
 %%%%%%%%%%%%%%%%%%%%%%%%%%%
 \section{Maxwell models of the Boltzmann equation}
 We consider a spatially homogeneous rarefied $d$-dimensional
 gas $(d = 2,3,\ldots)$ of particles having a unit mass.
 Let $f(v,t)$, where $v\in\real^d$ and $t\in\real_+$ denote respectively
 the velocity and time variables, be a  one-particle distribution
 function with the usual normalization
 \begin{equation}\label{eq2.1}
 \int_{\real^d} dv\, f(v,t) = 1\ .
 \end{equation}
 Then $f(v,t)$ has an obvious meaning of a time-dependent
 probability density in $\real^d$.
 We assume that the collision frequency is independent of the velocities
 of colliding particles (Maxwell-type interactions) and consider three
 different physical models  {\bf(A)},  {\bf (B)}  and  {\bf (C)} described below.

 {\bf (A)  Classical Maxwell gas (elastic collisions).}
 In this case  $f(v,t)$ satisfies the usual Boltzmann equation
 \begin{equation}\label{eq2.2}
f_t = Q(f,f)
 = \int_{\real^d \times S^{d-1}}
 \mkern-32mu   dw\,d\omega\, g
 \left( \frac{u\cdot \omega}{|u|} \right)
 \Big[f(v')f(w') - f(v)f(w)\Big]\ ,
 \end{equation}
 where the exchange of the velocities after a collision are given by
$$ v' = \frac12 (v+w+|u|\omega)\, ,\qquad \ \text{and}\ \ \quad
  w' = \frac12 (v+w - |u|\omega)
$$ where
 $u= v-w$ is the relative velocity and $\Omega \in S^{d-1}$.
 For the sake of brevity we shall consider below the model
 non-negative collision kernels $g(s)$ such that $g(s)$ is integrable
 on $[-1,1]$.
 The argument $t$  of $f(v,t)$ and similar functions is often omitted
 below (as in Eq.~\eqref{eq2.2}).

 {\bf (B) Elastic model with a thermostat}
 This case corresponds to  model  {\bf (A) }
in the presence of a thermostat  that consists of
 Maxwell particles with mass $m>0$  having the Maxwellian
 distribution
 \begin{equation}\label{eq2.3}
 M(v) = \left( \frac{2\pi T}{m}\right)^{-d/2}
 \exp \left( -\frac{m|v|^2}{2T}\right)
 \end{equation}
 with a constant temperature $T>0$.
 Then the evolution equation  for $f(v,t)$ becomes
 \begin{equation}\label{eq2.4}
 f_t = Q(f,f) +\theta \int dw\,d\omega\, g
 \left(\frac{u\cdot \omega}{|u|}\right)
 \Big[ f(v') M(w') - f(v) M(w)\Big]\ ,
 \end{equation}
 where  $\theta >0$ is a coupling constant, and the precollision
  velocities is now
$$
v' = \frac{v+m(w+|u|\omega)}{1+m},\qquad \ \text{and}\ \ \quad
 w' = \frac{v+mw - |u|\omega}{1+m},
$$
with  $u= v-w$ the relative velocity and
 $\omega \in S^{d-1}$.

 Equation~\eqref{eq2.4} was derived in \cite{BG-06} as a
 certain limiting case of a binary mixture of weakly interacting
 Maxwell gases.

{\bf (C)  Maxwell model for inelastic particles.}
 We consider this model in the form given in \cite{BC-03}.
 Then the inelastic Boltzmann equation in the weak form reads
 \begin{equation}\label{eq2.5}
 \frac{\partial}{\partial t} (f,\psi)
 = \int_{\real^d\times \real^d \times S^{d-1}} \mkern-36mu
 dv\, dw\, d\omega\, f(v) f(w)
 \frac{|u\cdot\omega|} {|u|}
 [\psi (v')-\psi (v)]\ ,
 \end{equation}
 where $\psi (v)$ is a bounded and continuous test function,
 \begin{equation}\label{eq2.6}
 (f,\psi) = \int_{\real^d} \mkern-12mu dv\, f(v,t)\psi (v),\
 u= v-w,\ \omega \in S^{d-1}, \
 v' = v-\frac{1+e}{2} (u\cdot \omega) \omega\ ,
 \end{equation}
 the constant parameter $0 < e\le 1$ denotes the restitution coefficient.
 Note that the model {\bf (C}  with $e=1$ is equivalent to the model {\bf (A)} with some kernel $g(s)$.

\medskip

 All three models can be simplified (in the mathematical sense) by
 taking the Fourier transform.
 We denote
 \begin{equation}\label{eq2.7}
 \hat f (k,t) = \F [f] = (f, e^{-ik\cdot v}),\qquad k\in\real^d\ ,
 \end{equation}
 and obtain (by using the same trick as in \cite{Bo76} for the model
 {\bf (A)})  for all three models the following equations:
 \begin{equation}\label{eq2.8}
 \text{\bf (A)}\quad\quad\quad\quad\quad\quad \quad
 \hat f_t = \widehat Q (\hat f,\hat f)
 = \int_{S^{d-1}} \mkern-20mu d\omega\, g\Big(\frac{k\cdot\omega}{|k|}\Big)
 [\hat f(k_+) \hat f(k_-) - \hat f(k) \hat f(0)]\ ,\quad\quad\quad\quad
 \end{equation}
 where
 $k_\pm = \frac12 (k\pm |k|\omega)$, $ \omega \in S^{d-1}$,
$ \hat f(0)=1$.
 \begin{equation}\label{eq2.9}
 \text{\bf (B)}\quad \quad\quad\quad\quad\quad
 \hat f_t = \widehat Q (\hat f,\hat f) + \theta \int_{S^{d-1}}\mkern-20mu
 d\omega\, g\Big( \frac{k\cdot\omega}{|k|}\Big)
 [\hat f(k_+) \widehat M (k_-) - \hat f(k) \widehat M(0)]\ ,\quad\quad\quad
 \end{equation}
 where
 $\widehat M (k) = e^{-\frac{T|k|^2}{2m}}$,
 $k_+ = \frac{k+m|k|\omega}{1+m}$,
 $k_- = k-k_+$,
$ \omega \in S^{d-1}$,
 $\hat f(0)=1$.
 \begin{equation}\label{eq2.10}
 \text{\bf (C)}\quad\quad\quad\quad\quad\quad \quad\quad\quad\quad\quad
 \hat f_t = \int_{S^{d-1}}\mkern-20mu
 d\omega \frac{|k\cdot\omega|}{|k|}
 [\hat f(k_+) \hat f(k_-) - \hat f (k) \hat f(0)]\ ,\quad\quad\quad\quad
\quad \quad\end{equation}
 where
 $k_+ = \frac{1+e}2 (k\cdot \omega)\omega$,
 $k_- = k-k_+$,
 $\omega \in S^{d-1}$,
 $\hat f (0) =1$. An equivalent formulation is given for $k_=
 \frac{1+e}{4}(k-|k|\omega),$ and
 $k_+ = k-k_-$ .

 The case {\bf (B)} can be obviously simplified by the  substitution
 \begin{equation}\label{eq2.11}
 \hat f(k,t) = \widetilde{\widehat  f} (k,t) \exp \left[ -
\frac{T|k|^2}2\right]\ .
 \end{equation}
 Then we obtain, omitting tildes in the final equation:
 \begin{equation}\label{eq2.12}
 \text{\bf (B$'$)}\quad \quad\quad\quad \quad\quad\quad
 \hat f_t = \widehat Q (\hat f,\hat f) + \theta \int_{S^{d-1}}\mkern-24mu
 d\omega\, g \Big( \frac{k\cdot\omega}{|k|}\Big)
 \left[ \hat f \Big( \frac{k+m|k|\omega}{1+m}\Big) - \hat f(k)\right]\ ,
\quad\quad \quad
 \end{equation}
 i.e., the  case {\bf (B)} with $T=0$.
 Therefore, we shall consider below just the case {\bf (B$'$)},
 assuming nevertheless
 that $\hat f(k,t)$ in Eq.~\eqref{eq2.12} is the Fourier transform
 \eqref{eq2.7}
 of a probability density $f(v,t)$.

%%  that solves Eq.
%%  \eqref{eq2.4} after the convolution with the inverse Fourier
%% transform of $ \exp ({-T|k|^2/}2)$.

%%%%%%%%%%%%%%%%%%%%%%%%%%%%%%%%%%%%%%%%%%%%%%%%%%%%%%%%%%%%%%%%%%
 %%%%%%%%%%%%%%%%%%%% section 3  %%%%%%%%%%%%%%%%%%%%%%%%%%%%%%%%%%%
%%%%%%%%%%%%%%%%%%%%%%%%%%%%%%%%%%%%%%%%%%%%%%%%%%%%%%%%%%%%%
\vskip.5in

 \section{Isotropic Maxwell model in the Fourier representation}

 We shall see that the three models {\bf  (A), (B)} and {\bf  (C)}
 admit a class of isotropic solutions with
 distribution functions $f= f(|v|,t)$. Then, according to   \eqref{eq2.7}
 we look for solutions $\hat f= \hat f(|k|,t)$
 to  the  corresponding isotropic Fourier transformed
 problem, given by
 \begin{equation}\label{eq3.1}
 x= |k|^2\ ,\quad
 \varphi (x,t) = \hat f(|k|,t) = \F[f(|v|,t)]\ ,
 \end{equation}
 where  $\varphi (x,t)$ solves the following initial value problem
 \begin{equation}\label{eq3.2}
 \begin{split}
 \varphi_t & = \int_0^1 ds\, G(s) \left\{ \varphi [a(s)x] \varphi [b(s)x]
 - \varphi (x)\right\} +\\
 &\qquad +
 \int_0^1 ds H(s) \left\{ \varphi [c(s)x] - \varphi (x)\right\}\ ,\\
&\ \text{}
\quad \varphi_{{t=0}} =\varphi_{{0}}(x)\, ,\  \ \  \text{}\ \  \varphi (0,t)=1\ ,
 \end{split}
 \end{equation}
 where $a(s)$, $b(s)$, $c(s)$ are non-negative continuous functions  on
 $[0,1]$, whereas $G(s)$ and $H(s)$ are generalized non-negative
 functions such that
 \begin{equation}\label{eq3.3}
 \int_0^1 ds\, G(s) <\infty\ ,\qquad
 \int_0^1 ds\, H(s) < \infty\ .
 \end{equation}
 Thus, we do not exclude such functions as $G=\delta (s-s_0)$,
 $0< s_0 <1$, etc.
 We shall see below that, for isotropic solutions \eqref{eq3.1}, each
 of the three equations \eqref{eq2.8}, \eqref{eq2.10}, \eqref{eq2.12}
 is a particular case of Eq.~\eqref{eq3.2}.

 Let us first consider Eq.~\eqref{eq2.8} with $\hat f (k,t) = \varphi (x,t)$
 in the notation \eqref{eq3.1}.
 In that case
 $$|k_\pm|^2 = |k|^2 \frac{1\pm (\omega_0 \cdot\omega)}{2}\ ,\qquad
 \omega_0 = \frac{k}{|k|} \in S^{d-1}\ ,\  d=2,\ldots,$$
 and the integral in Eq.~\eqref{eq2.8} reads
 \begin{equation}\label{eq3.4}
 \int_{S^{d-1}} d\omega\, g (\omega_0\cdot \omega) \varphi
 \left[ x\frac{1+\omega_0 \cdot \omega}{2} \right] \varphi
 \left[ x\frac{1-\omega_0\cdot\omega}{2}\right]\  .
 \end{equation}
 It is easy to verify the identity
 \begin{equation}\label{eq3.5}
 \int_{S^{d-1}} d\omega\, F (\omega\cdot\omega_0)
 = |S^{d-2}| \int_{-1}^1 dz\, F(z) (1-z^2)^{\frac{d-3}2}\ ,
 \end{equation}
 where $|S^{d-2}|$ denotes the ``area'' of the unit sphere in $\real^{d-1}$
 for $d\ge 3$ and $|S^0 | =2$.
 The identity \eqref{eq3.5} holds for any function $F(z)$ provided the
 integral as defined  in the right hand side of \eqref{eq3.5} exists.

 The integral \eqref{eq3.4} now reads
 \begin{equation*}
 \begin{split}
 & |S^{d-2}| \int_{-1}^1 dz\, g (z) (1-z^2)^{\frac{d-3}2}
 \varphi \Big( x\frac{1+z}2\Big) \varphi \Big( x\frac{1-z}2\Big) =\\
 &\qquad = \int_0^1 ds\, G(s)\varphi (sx) \varphi [(1-s)x]\ ,
 \end{split}
 \end{equation*}
 where
 \begin{equation}\label{eq3.6}
 G(s) = 2^{d-2} |S^{d-2}| g(1-2s) [s(1-s)]^{\frac{d-3}2}\ ,\qquad
 d= 2,3,\ldots\ .
 \end{equation}

 Hence, in this case we obtain Eq.~\eqref{eq3.2}, where
 \begin{equation}\label{eq3.7}
 \text{\bf (A)}\qquad \qquad \qquad  \qquad
 a(s) =s\ ,\qquad b(s) = 1-s\ ,\qquad H(s)=0\ ,\quad\quad \quad
\qquad \quad\quad \quad
 \end{equation}
 $G(s)$ is given in Eq.~\eqref{eq3.6}.

 Two other models  {\bf (B$'$)} and {\bf (C)}, described by
Eqs.~\eqref{eq2.12},
 \eqref{eq2.10} respectively, can be considered quite similarly.
 In both cases we obtain Eq.~\eqref{eq3.2}, where
 \begin{equation}\label{eq3.8}
 \begin{split}
 \text{\bf (B$'$)}\qquad \qquad \qquad  \qquad  \qquad
 &a(s) =s,\quad
 b(s) = 1-s,\quad
 c(s) = 1- \frac{4m}{(1+m)^2} s,\qquad \quad\quad\quad \\
 &H(s) = \theta G(s)\ ,
 \end{split}
 \end{equation}
 $G(s)$ is given in Eq.~\eqref{eq3.6}; while for the inelastic
 collision case
 \begin{equation}\label{eq3.9}
 \begin{split}
 \text{{\bf (C)}} \qquad \qquad \qquad \qquad
& a(s) = \frac{(1+e)^2}4 s,\quad
 b(s) =1 - \frac{(1+e)(3-e)}{4}\, s,\qquad\qquad \quad\quad\quad \\
 \noalign{\vskip6pt}
& H (s)=0,\quad
 G(s) = |S^{d-2}| (1-s)^{\frac{d-3}2}\ .
\end{split}
 \end{equation}

 Hence, all three models are described by Eq.~\eqref{eq3.2} where
 $a(s) \le 1$, $b(s) \le 1$, $c(s)\le 1$ are non-negative linear functions.
 One can also find in recent publications some other  useful equations that
 can be reduced after Fourier or Laplace transformations to
 Eq.~\eqref{eq3.2} (see, for example, \cite{BN05},
 \cite{PT05} that correspond to the case
 $G= \delta (s-s_0)$, $H=0$).

 The Eq.~\eqref{eq3.2} with $H(s) =0$ first appeared in its general
 form in the paper \cite{BC-03} in connection with models {\bf (A)}  and {\bf
(C)}.
 The consideration of the problem of self-similar asymptotics for
 Eq.~\eqref{eq3.2} in that paper made it quite clear that the most
 important properties of ``physical'' solutions depend very weakly on
the  specific functions $G(s)$, $a(s)$ and $b(s)$.
% Our goal in this paper is to study Eq.~\eqref{eq3.2} under very
% general restrictions on functions $a,b,c,G,H$.
% In fact, as we shall see below, these properties (large time asymptotics)
% are  typical for a much larger class of equations with nonlinearities
% of any order.

\vskip.5in
%%%%%%%%%%%%%%%%%%%%%%%%%%%%%%%%%%%%%%%%%%%%%
 %%%%%%%%%%%%%%%%%%  section 4 %%%%%%%%%%%%%%%%%%%%%
%%%%%%%%%%%%%%%%%%%%%%%%%%%%%%%%%%%%%%%
 \section{Models with multiple interactions and\\ statement of
 the general problem}

We shall present in this section a general framework to study  solutions
to  the type of problems introduced in the previous section.

 We assume without loss of generality (scaling transformations
 $\tilde t = \alpha t$, $\alpha = \text{const.}$) that
 \begin{equation}\label{eq4.1}
 \int_0^1 ds\, [G(s) + H(s)] = 1
 \end{equation}
 in Eq.~\eqref{eq3.2}.
 Then Eq.~\eqref{eq3.2} can be considered as a particular case
 of the following equation for a function $u(x,t)$
 \begin{equation}\label{eq4.2}
 u_t + u = \Gamma (u)\ ,\qquad x\ge 0,\  t\ge 0\ ,
 \end{equation}
 where
 \begin{equation}\label{eq4.3}
 \begin{split}
 &\Gamma (u) = \sum_{n=1}^N \alpha_n \Gamma^{(n)} (u)\ ,\quad
 \sum_{n=1}^N \alpha_n =1\ ,\ \alpha_n \ge 0\ ,\\
& \Gamma^{(n)} (u) = \int_0^\infty\mkern-20mu  da_1  \ldots
\int_0^\infty \mkern-20mu da_n\
 A_n (a_1,\ldots, a_n) \prod_{k=1}^n  u(a_kx),\quad
 n=1,\ldots,N\ .
 \end{split}
 \end{equation}
 We assume that
 \begin{equation}\label{eq4.3bis}
A_n(a) = A_n (a_1,\ldots, a_n) \ge 0\ ,\qquad
 \int_0^\infty da_1 \ldots \int_0^\infty da_n\  A(a_1,\ldots,a_n) =1\ ,
 \end{equation}
 where  $A_n(a) = A_n (a_1,\ldots, a_n)$ is a
 generalized density of a probability
 measure in $\real_+^n$ for any $n=1,\ldots,N$.
 We also assume that all $A_n(a)$ have a compact support, i.e.,
 \begin{equation}\label{eq4.4}
 A_n(a_1,\ldots,a_n) \equiv 0\ \text{ if }\
 \sum_{k=1}^n a_k^2 > R^2\ ,\qquad
 n = 1,\ldots, N\ ,
 \end{equation}
 for sufficiently large $0< R<\infty$.

 Eq.~\eqref{eq3.2} is a particular case of Eq.~\eqref{eq4.2} with
 \begin{equation}\label{eq4.5}
 \begin{split}
 &N= 2\ ,\quad \alpha_1 = \int_0^1 ds\, H(s)\ ,\quad
 \alpha_2 = \int_0^1 ds\, G(s)\\
 &A_1 (a_1) = \frac1{\alpha_1} \int_0^1 ds\, H(s) \delta [a_1 - c(s)]\\
 &A_2 (a_1,a_2) = \frac1{\alpha_2} \int_0^1 ds\, G(s)
 \delta [a_1 -a(s)] \delta [a_2-b(s)]\ .
 \end{split}
 \end{equation}

 It is clear that Eq.~\eqref{eq4.2} can be considered as a
 generalized Fourier transformed isotropic Maxwell model with
 multiple interactions provided $u(0,t) =1$, the case $N=\infty$
 in Eqs.~\eqref{eq4.3} can be treated in the same way.

\vskip.5in

 The general problem  we consider below can be formulated
 in the following way.
 We study the initial value problem
 \begin{equation}\label{eq4.6}
 u_t + u = \Gamma (u)\ ,\quad
 u_{|t=0} = u_0 (x)\ ,\quad
 x\ge 0\ ,\quad
 t\ge 0\ ,
 \end{equation}
 in the Banach space $B= C(\real_+)$ of continuous functions $u(x)$
 with the norm
 \begin{equation}\label{eq4.7}
 \| u\| = \sup_{x\ge 0} |u(x)|\ .
 \end{equation}

 It is usually assumed that $\|u_0\| \le 1$
and that the operator
 $\Gamma$ is given by Eqs. ~\eqref{eq4.3}.
 On the other hand, there are just a few properties of $\Gamma (u)$
 that are essential for existence, uniqueness and large time asymptotics
 of the solution $u(x,t)$ of the problem~\eqref{eq4.6}.
Therefore, in many cases  the results can be applied to more
general classes of operators $\Gamma$ in Eqs.~\eqref{eq4.6} and
more general functional space, for example $B= C(\real^d)$
(anisotropic models).
That is why we study below the class \eqref{eq4.3} of operators
$\Gamma$ as the most important example, but simultaneously indicate which
properties of $\Gamma$ are relevant in each case.
In particular, most of the  results of Section~4--7 do not use a specific form
\eqref{eq4.3} of $\Gamma$ and, in fact, are valid for a  more general class of
operators.

Following this way of study, we first consider the problem \eqref{eq4.6}
with $\Gamma$ given by Eqs.~\eqref{eq4.3} and point out the most
important properties of $\Gamma$.

We simplify notations and omit in most of the cases below the argument
$x$ of the function $u(x,t)$.
The notation $u(t)$ (instead of $u(x,t)$) means then the function of the  real
variable $t\ge 0$ with values in the space $B = C(\real_+)$.

\begin{remark}
We shall omit below the argument $x\in \real_+$ of functions $u(x)$, $v(x)$,
etc., in all cases when this does not cause a misunderstanding.
In particular, inequalities of the kind $|u| \le |v|$ should be
understood as a  point-wise  control in absolute value, i.e.
 ``$|u(x)| \le |v(x)|$ for any $x\ge 0$'' and so on.
\end{remark}

We first start by giving the following general definition for
 operators  acting on a unit ball  of a Banach space $B$ denoted by
\begin{equation}\label{eq4.10}
U = \{u \in B : \|u\| \le 1\}
\end{equation}

\medskip

\begin{definition}\label{def4.1} The operator  $\Gamma=\Gamma(u)$
is called an  $L$-Lipschitz operator if there exists a linear bounded
operator $L: B\to B$  such that
the inequality
\begin{equation}\label{eq5.5}
| \Gamma (u_1) - \Gamma (u_2)| \le L (|u_1 - u_2 |)
\end{equation}
holds for any pair of functions $ u_{1,2}$ in  $U $.
\end{definition}
\begin{remark}
Note that the $L$-Lipschitz condition  \eqref{eq5.5} holds, by definition,
at any point $x\in \real_+$ (or  $x\in \real^d$ if $B = C(\real^d)$).
Thus, condition
 \eqref{eq5.5} is much stronger than the classical Lipschitz condition
\begin{equation}\label{eq4.13}
\| \Gamma (u_1) - \Gamma (u_2) \|
< C \| u_1 - u_2\| \ \text{ if }\ \ u_{1,2} \in U
\end{equation}
which obviously follows from  \eqref{eq5.5} with the constant $C=\|L\|_B$,
the norm of the operator $L$ in the  space of bounded operators acting in $B$.
In other words, the terminology ``$L$-Lipschitz condition'' means the
point-wise Lipschitz condition with respect to an specific linear operator $L$.
\end{remark}
\medskip

The next lemma shows that the operator   $\Gamma(u)$ defined in
 Eqs.~\eqref{eq4.3}, which  satisfies $\Gamma(1)=1$ (mass conservation)
and  maps $U$ into itself,
 satisfies an   $L$-Lipschitz condition, where
 the
linear operator $L$ is the one given by the linearization of $\Gamma$
near  the unity.

 We assume without loss of generality that the kernels
$A_n (a_1,\ldots,a_n)$ in Eqs.~\eqref{eq4.3}
are symmetric with respect to any permutation of the arguments
$(a_1,\ldots,a_n)$, $n=2,3,\ldots,N$.

\begin{thm}\label{thm-lip} The operator   $\Gamma(u)$ defined in
 Eqs.~\eqref{eq4.3} maps $U$ into itself  and
satisfies the $L$-Lipschitz condition \eqref{eq5.5},
where the linear operator  $L$ is given by
\begin{equation}\label{eq5.2}
Lu = \int_0^\infty da\, K (a) u (ax)\ ,
\end{equation}{with}
\begin{equation}\begin{split}\label{eq5.3}
K(a) = \sum_{n=1}^N n\alpha_n K_n (a)\ ,\\
\text{where}\qquad
K_n(a) = \int_0^\infty da_2\ldots \int_0^\infty  da_n \ A_n
(a_1,a_2,\ldots,a_n) \ \ \text{and}\  \ \sum_{n=1}^N \alpha_n =1\, .
\end{split}\end{equation}
for symmetric kernels $A_n(a_1,a_2,\ldots,a_n), n=2,\ldots$\ .
\end{thm}

\begin{proof}
First, the operator   $\Gamma(u)$ in \eqref{eq4.3}-\eqref{eq4.4}
 maps $B$ to itself and also satisfies
\begin{equation}\label{eq4.8}
\|\Gamma (u)\| \le \sum_{n=1}^N \alpha_n \|u\|^n\ ,\qquad
\sum_{n=1}^N \alpha_n =1\ .
\end{equation}
Hence,
\begin{equation}\label{eq4.9}
\|\Gamma (u)\| \le 1\ \text{ if }\ \|u\| \le 1\ ,
\end{equation}
and then  $\Gamma (U) \subset U$, so its maps $U$ into itself.

Since $\Gamma (1) =1$, we introduce the linearized operator
$L:B\to B$ such that formally
\begin{equation}\label{eq5.1}
\Gamma (1+\ep u) = 1+\ep Lu + O(\ep^2)\ .
\end{equation}
By using the symmetry of kernels $A_n(a)$, $n=2,3,\ldots,N$,
one can easily check that
\begin{equation*}
Lu = \int_0^\infty da\, K (a) u (ax)\ ,
\end{equation*}
where
\begin{equation*}\label{}
K(a) = \sum_{n=1}^N n\alpha_n K_n (a)\ ,\quad
K_n(a) = \int_0^\infty da_2\ldots \int_0^\infty da_n \ A_n
(a,a_2,\ldots,a_n)\ .
\end{equation*}
Note that $K(a) \ge 0$ has a compact support and
\begin{equation*}\label{}
\int_0^\infty da\, K(a) = \sum_{n=1}^N n\alpha_n\ ,\qquad
\sum_{n=1}^N \alpha_n =1\ ,
\end{equation*}
because of the conditions \eqref{eq4.3}, \eqref{eq4.4}. So conditions
\eqref{eq5.2}-\eqref{eq5.3} are satisfied.

In addition, in order to proof the $L$-Lipschitz
property~\eqref{eq5.5} for the operator $\Gamma$ given in Eqs.
\eqref{eq4.3}, we make use of the multi-linear structure of the
integrand associated with the definition of $\Gamma(u)$. Indeed,
from the elementary identity
$$ab - cd = \frac{a+c}2 (b-d) + \frac{b+d}2 (a-c)\, ,$$
we estimate
\begin{equation*}
\begin{split}
&|u_1 (a_1x) u_1 (a_2x) - u_2 (a_1x) u_2(a_2x)| \\
&\qquad  \le |u_1 (a_1x) - u_2(a_1x)|  +
  |u_1(a_2x) - u_2(a_2x)|\ ,\qquad x\ge 0\ ,\ a_{1,2} \ge 0\ ,
\end{split}
\end{equation*}
provided $\|u_{1,2}\| \le 1$.
Then we obtain
$$|\Gamma^{(2)} (u_1) - \Gamma^{(2)}(u_2)|
\le 2 \int_0^\infty da\, K_2(a) | u_1(ax) - u_2(ax)|$$
in the notation of Eqs.~\eqref{eq4.3}, \eqref{eq5.3}.
It remains to prove that
\begin{equation}\label{eq5.6}
|\Gamma^{(n)} (u_1) - \Gamma^{(n)} (u_2)|
\le n \int_0^\infty da\, K_n (a) |u_1(ax) - u_2 (ax)|
\end{equation}
for $3\le n\le N$ (the case $n=1$ is trivial).
This problem can be obviously reduced to an elementary inequality
\begin{equation}\label{eq5.7}
\Big| \prod_{k=1}^n x_k - \prod_{k=1}^n y_k\Big|
\le \sum_{k=1}^n |x_k - y_k|\ ,\qquad
n=3,\ldots,
\end{equation}
provided $|x_k| \le 1$, $|y_k| \le 1$, $k=1,\ldots,n$.
Since this is true for $n=2$, we can use the induction.
Let
$$a= x_{n+1} \ ,\quad
c= y_{n+1}\ ,\quad
b = \prod_{k=1}^n x_k\ ,\quad
d = \prod_{k=1}^n y_k\ ,$$
then
\begin{equation*}
\begin{split}
\Big| \prod_{k=1}^{n+1} x_k - \prod_{k=1}^{n+1} y_k\Big|
& = |ab-cd| \le |a-c| + |b-d|\le\\
& \le |x_{n+1} - y_{n+1}|
+ \sum_{k=1}^n |x_k - y_k|\ ,
\end{split}
\end{equation*}
and the inequality \eqref{eq5.7} is proved for any $n\ge 3$.
Then we proceed exactly as in case $n=2$ and prove the estimate
\eqref{eq5.7} for arbitrary $n\ge 3$.
Inequality \eqref{eq5.5} follows  directly from the
definition of operators $\Gamma$ and $L$.
\end{proof}

\begin{cor}             %\label{cor5.2}
The Lipschitz condition \eqref{eq4.13} is fulfilled for $\Gamma (u)$
given in Eqs.~\eqref{eq4.3} with the constant
\begin{equation}\label{eq5.8}
C = \|L\| = \sum_{n=1}^N n\alpha_n\ ,\qquad
\sum_{n=1}^\infty \alpha_n =1\ ,
\end{equation}
where $\|L\|$ is the norm of $L$ in $B$.
\end{cor}

\begin{proof}
The proof follows directly from the inequality \eqref{eq5.5} and
Eqs.~\eqref{eq5.2}, \eqref{eq5.3}.
\end{proof}

 It is also easy to prove that the $L$-Lipschitz condition holds in
 $B=C(\real^d)$ for ``gain-operators'' in the  Fourier transformed Boltzmann
equations  ~\eqref{eq2.8},  \eqref{eq2.9} and \eqref{eq2.10}.

%%%%%%%%%%%%%%%%%%%%%%%%%%%%%%%%%%%%%%%%%%%%%%%%%%%%%%%%
%%%%%%%%%%%%%%%%% Section 5 %%%%%%%%%%%%%%%%%%%%%%%%%%
%%%%%%%%%%%%%%%%%%%%%%%%%%%%%%%%%%%%%%%%%%%%%%%%%%%%

\vskip.5in
\section{Existence and uniqueness of solutions}

The aim of this  section is to state and prove, with
 minimal requirements,  the existence and uniqueness
results associated with the initial value problem \eqref{eq4.6} in the Banach
 space $(B,\| \cdot \|)$,
 where the norm associated to B is defined in \eqref{eq4.7}
 In fact, this existence and uniqueness result is an
 application of the classical Picard iteration scheme
 and holds for any  operator $\Gamma$
 which satisfies  the usual Lipschitz condition
 \eqref{eq4.13} and  transforms
 the unit ball $U$ into itself.
We include its proof for the sake of completeness.

\begin{lem}\label{lem5.1} ({\sl Picard Iteration scheme})
If the conditions in \eqref{eq4.9} and \eqref{eq4.13} are fulfilled
then the initial value
problem \eqref{eq4.6} with arbitrary $u_0 \in U$ has a
unique solution $u(t)$ such that $u(t) \in U$ for any $t\ge 0$.
\end{lem}

\begin{proof}
Consider   the integral form of Eq.~\eqref{eq4.2}
\begin{equation} \label{eq4.11}
u(t) = u_0 e^{-t} + \int_0^t d\tau\, e^{-(t-\tau)} \Gamma [u(\tau)]
\end{equation}
and apply the standard Picard iteration scheme
\begin{equation}\label{eq4.12}
u^{(n+1)} (t) = u_0 e^{-t} + \int_0^t d\tau\, e^{-(t-\tau)} \Gamma
[u^{(n)} (\tau)],\quad
u^{(0)} = u_0\ .
\end{equation}
Consider a finite interval $0\le t\le T$ and denote
$$\Norm{u}_T = \sup_{0\le t\le T} \|u(t)\|\ .$$
Then
$$\| u^{(n+1)} (t)\| \le \|u_0\| e^{-t} + (1-e^{-t}) \Norm{\Gamma (u^{(n)})}_t$$
and therefore,  by induction
$$\|u^{(n)}(t)\| \le 1\ \text{ for all }\ n=1,2,\ldots, \ \text{ and }\ t\in [0,T]\ ,$$
since $\|u_0\| \le 1$ and $\Gamma (u)$ satisfies the inequality \eqref{eq4.9}.
If, in addition,  $\Gamma (u)$ satisfies the Lipschitz condition
\eqref{eq4.13},
then it is easy to verify that
$$\Norm{u_{n+1} - u_n}_T \le (1-e^{-T}) C\Norm{u_n - u_{n-1}}_T\ ,
\qquad n= 1,2,\ldots,$$
and therefore, the iteration scheme \eqref{eq4.12} converges
uniformly for any $0\le t\le T$ provided
\begin{equation}\label{eq4.14}
C(1-e^{-T}) < 1\Longrightarrow T < \ln \frac{C}{C-1} \ \  \text{ if }\
C > 1
\end{equation}
($T$ can be taken arbitrary large if $C\le 1$).
It is easy to verify that
$$u(t) = \lim_{n\to\infty} u^{(n)} (t)\ ,\qquad 0\le t\le T\ ,$$
is a solution of Eqs.~\eqref{eq4.6}, \eqref{eq4.11}, satisfying the
inequality
\begin{equation}\label{eq4.15}
\|u (t) \| \le 1\ ,\qquad 0\le t\le T\ .
\end{equation}
The Lipschitz condition \eqref{eq4.13} is also sufficient to show that
this solution is unique in a class of functions satisfying the inequality
\eqref{eq4.15} on any interval $0\le t\le \ep$.

This proof of existence and uniqueness for the
Cauchy problem \eqref{eq4.6}-\eqref{eq4.7} is quite standard (see any textbook
on ODEs) and therefore we omit some details.

The important point is that in our case the length $T$ of the
initial time-interval does not depend  on the initial conditions
(see Eq.~\eqref{eq4.14}).
Therefore, we can proceed by considering the interval $T\le t\le 2T$
and so on.
Thus we obtain the global in time solution $u(t) \in U$, where $U$ is
the closed unit ball in $B$, of the Cauchy problem \eqref{eq4.6}.
\end{proof}

 \begin{thm}\label{thm5.2}
 Consider the Cauchy problem \eqref{eq4.6} with $\|u_0\| \le1$
 and assume that the operator $\Gamma :B\to B$
 \begin{itemize}
 \item[{\bf (a)}]
 maps the closed unit ball $U\subset B$ to itself, and
\item[{\bf (b)}] satisfies  a  $L$-Lipschitz
condition  \eqref{eq5.5}  for some
 positive bounded linear operator $L:B\to B$.
 \end{itemize}
 Then
\begin{itemize}
\item[{\bf i)}]
there exists a unique solution $u(t)$ of the problem
\eqref{eq4.6} such that ${\|u(t)\| \le1}$ for any $t\ge 0$;
\item[{\bf ii)}]
any two solutions $u(t)$ and $w(t)$  of problem \eqref{eq4.6} with initial
data in  the unit ball $U$ satisfy the inequality
\begin{equation}\label{eq5.11}
|u(t) - w(t)| \le \exp\{t(L-1)\} (|u_0-w_0|)\ .
\end{equation}
\end{itemize}
\end{thm}

\begin{proof}
The proof of {\bf i)} follows directly from Lemma~\ref{lem5.1} (with the
Lipschitz constant $C = \|L\|$).
For the proof of {\bf ii)},
let $u(t)$ and $w(t)$ be two solutions of this problem such that
\begin{equation}\label{eq5.9}
u(0) = u_0\ ,\quad
w(0) = w_0\ ,\quad
\|u_0\| \le 1\ ,\quad
\|w_0\| \le 1\ .
\end{equation}
Then the function $y(t) = u(t) - w(t)$ satisfies
the equation
$$ y_t + y = g(x,t) = \Gamma (u) - \Gamma (w)\ ,
\qquad y|_{ t=0} = u_0 - w_0 = y_0\ .$$
Hence,
$$y(t) = e^{-t} y_0 + \int_0^t d\tau\, e^{-(t-\tau)} g(x,t)\ ,$$
and, applying theorem~\ref{thm-lip} for the use of  the $L$-Lipschitz
condition  \eqref{eq5.5} , we obtain
\begin{equation}\label{eq5.10}
|y(t)| \le |y_0| e^{-t} + \int_0^t d\tau\, e^{-(t-\tau)} L|y(\tau)|\ .
\end{equation}
Clearly,  $|y(t)|\le y_*(t)$, where $y_*(t)$ satisfies
the equation
\begin{equation}\label{eq5.10.1}
y_* (t) = |y_0| e^{-t} + \int_0^t d\tau\, e^{-(t-\tau)} Ly_*(\tau)\ .
\end{equation}
Since, by  {\bf (b)}, the linear operator $L:B\to B$ is positive and bounded,
then  Eq.~\eqref{eq5.10.1} has a unique solution
$$y_*(t) = e^{-t(1-L)} |y_0| = e^{-t} \sum_{n=0}^\infty
\frac{t^n}{n!} L^n |y_0|$$
so  the  estimate \eqref{eq5.11} follows and the proof of the
theorem is completed.
\end{proof}

Theorem~\ref{thm-lip} and the inequality \eqref{eq4.8} show that the
operator $\Gamma$ given in Eqs.~\eqref{eq4.3} satisfies all conditions
of the theorem.

\begin{remark}
{\bf $1 -$} The above consideration is, of course, a simple
generalization of the proof of the usual Gronwall inequality for the
scalar function $y(t)$. The essential difference is, however, that
$y(t)$ is a ``vector'' $y(x,t)$ with values in the Banach space $B=
C(\real_+)$ and, consequently, the estimate \eqref{eq5.11} for the
functions $u(x,t)$ and $w(x,t)$ holds at any point $x\in \real_+$.

{\bf $2 -$} We stress that  estimates
\eqref{eq5.10}-\eqref{eq5.10.1} do not depend on specific
properties of the operator $L$ beyond that  of being positive and bounded.
The Banach space $B = C(\real_+)$ can be also replaced,
for example, by $B= C(\real^d)$ (that is the case of
non-isotropic models).
Therefore we formulated Theorem~\ref{thm5.2} in the most general form
(without specifying the particular space $B$ of continuous functions).
\end{remark}

We remind the reader that the initial value problem \eqref{eq4.6}
appeared as a generalization of the initial value problem
\eqref{eq3.2} for a
characteristic function  $\varphi (x,t)$, i.e., for the Fourier
transform  of a  probability measure (see
Eqs.~\eqref{eq3.1}, \eqref{eq2.1}).
It is important therefore to show that the solution $u(x,t)$ of the
problem \eqref{eq4.6} is a characteristic function for any $t>0$
provided this is so for $t=0$.
The answer to this and similar questions is given in the
following statement.

\begin{lem}\label{lem5.3}
Let $U' \subset U\subset B$ be any closed convex subset of the
unit ball $U$ (i.e., $u = (1-\theta) u_1 + \theta u_2 \in U'$ for any
$u_{1,2} \in U'$ and $\theta\in [0,1]$).
If $u_0 \in U'$ in Eq.~\eqref{eq4.6} and $U$ is replaced by $U'$
in the condition~(1) of Theorem~\ref{thm5.2}, the theorem holds
and $u(t) \in U'$ for any $t\ge 0$.
\end{lem}

\begin{proof}
The only important point which should be changed in the proof is
the consideration of Eqs.~\eqref{eq4.12} in the proof of
Lemma~\ref{lem5.1}.
We need to verify that, for any $u_0\in U'$ and $v(\tau)\in U'$,
$0\le \tau\le t$,
\begin{equation}\label{eq5.12}
\hat u (t) = u_0 e^{-t} + \int_0^t d\tau\, e^{-(t-\tau)} v(\tau) \in U'\ .
\end{equation}
 In order to see that \eqref{eq5.12} holds,
 we note that $v(\tau) = \Gamma [u^{(n)}(\tau)]$ in
 Eqs.~\eqref{eq4.12} is, by construction, a continuous function of
 $\tau \in [0,t]$ and that
 $$e^{-t} + \int_0^t d\tau\, e^{-(t-\tau)} =1\ .$$

 Therefore we can approximate $\hat u (t)$ by an integral sum
 $$\hat u_m (t) = u_0 e^{-t} + \sum_{k=1}^m \gamma_k(t) v_k(\tau_k)\ ,
 \qquad \sum_{k=1}^m \gamma_k (t) = 1- e^{-t}\ ,$$
Then $u_m (t) \in U'$ as a convex linear combination of elements
of $U'$.
Taking the limit $m\to\infty$ ($U'$ is a  closed subset of $U$)
it follows that $\hat u(t)\in U'$ so  \eqref{eq5.12} holds.
Hence, the corresponding sequence
as in Eqs.~\eqref{eq4.12}  also satisfies
 $u^{(n)} (t)\in U'$  for all $n\ge 0$.
The rest of the proof continues as
the one of   Lemma~\ref{lem5.1}, so that
the result remains true for any closed convex
subset $U'\subset U$ and so does Theorem~\ref{thm5.2}.
Thus the proof of Lemma~\ref{lem5.3} is completed.
\end{proof}

\begin{remark} It is well-known (see, for example, the textbook
\cite{Fe}) that the
set $U'\subset U$ of Fourier transforms of probability measures
in $\real^d$ (Laplace transforms in the case of $\real_+$) is convex
and closed with respect to uniform convergence.
On the other hand, it is easy to verify that the inclusion $\Gamma(U')
\subset U'$, where $\Gamma$ is given in Eqs.~\eqref{eq4.3}, holds in
both cases of Fourier and Laplace transforms.
Hence, all results obtained for Eqs.~\eqref{eq4.2}, \eqref{eq4.3} can
be interpreted in terms of ``physical'' (positive and satisfying the
condition \eqref{eq2.1}) solutions of corresponding Boltzmann-like
equations with multi-linear structure of any order.

We also note that all results of this section remain valid for operators
$\Gamma$ satisfying conditions \eqref{eq4.3} with a  more general condition
such as
\begin{equation}\label{eq.5.10.bis}
\sum_{n=1}^N \alpha_n \le 1\ ,\qquad \alpha_n \ge 0\ ,
\end{equation}
so that  $\Gamma(1) <1 $ and so the mass may not be conserved.
The only difference in this case is that
the operator $L$  satisfying conditions \eqref{eq5.2}, \eqref{eq5.3}
is not a linearization
of    $\Gamma(u)$ near the unity. Nevertheless
Theorem~\ref{thm-lip} remains true.
The inequality \eqref{eq.5.10.bis} is typical for Fourier (Laplace)
transformed Smoluchowski-type equations where the
total number of particles is decreasing in time  (see \cite{MePe1, MePe2} for related work).
\end{remark}

%\vskip.5in
%%%%%%%%%%%%%%%%%%%%%%%%%%%%%%%%%%%%%%%%%%%%%%%%
%%%%%%%%%%%%%%%%%%%%%% section 6 %%%%%%%%%%%%%%%
%%%%%%%%%%%%%%%%%%%%%%%%%%%%%%%%%%%%%%%%%

\section{Large time asymptotics and self-similar solutions}

In this section we study in more detail the solutions to the initial
value   problem
\eqref{eq4.6}-\eqref{eq4.7} constructed in Theorem~\ref{thm5.2}
and,  in particular,
their long time behavior.

We  point out that
the long time asymptotics results
 are just a consequence of  some very general properties of
 operators
$\Gamma$ and its corresponding  $L$,
 namely, that $\Gamma$ maps the unit ball $U$ of the
Banach space
$B=C(\real_+)$ into itself,   $\Gamma$ is an  $L$-Lipschitz
operator (i.e. satisfies \eqref{eq5.5})  and that $\Gamma$ is invariant
under dilations.

These three properties are sufficient to study self-similar solutions
and large time asymptotic behavior for the solution to the Cauchy problem
\eqref{eq4.6} in the unit ball $U$ of the
Banach space $C(\real_+)$.

First of all, we show that the operator $\Gamma$ given in
Eqs.~\eqref{eq4.3} has these properties.

\bigskip

\vbox{
{\bf Main properties of the operator
$\Gamma$:}  \hfill
{\sl
 \begin{itemize}
\item[{\bf a)}] $\Gamma$ maps the unit ball $U$ of the Banach space
$B=C(\real_+)$ into itself, that is
\begin{equation}\label{eq6.1bis}
\qquad \|\Gamma (u)\| \le1\ \text{ for any } \ \ u\in C(\real_+)
\ \ \text{ such that} \ \
\|u\|\le 1.
\end{equation}
 \item[{\bf b)}]$\Gamma$ is an  $L$-Lipschitz
operator (i.e. satisfies \eqref{eq5.5}) with $L$ from \eqref{eq5.2}, i.e.
 \begin{equation}\label{eq6.1}
\qquad \  |\Gamma (u_1) - \Gamma (u_2)|(x)
\le L(|u_1-u_2|)(x) = \int_0^\infty da K (a) | u_1 (ax) - u_2 (ax)|\ ,\qquad
\end{equation}
for $K(a)\geq 0$, for all $x\ge 0$ and for any two functions $u_{1,2} \in C(\real_+)$ such that
$\|u_{1,2}\| \le 1$.
\item[{\bf c)}] $\Gamma$ is invariant
under dilations:
\begin{equation}\label{eq6.2}
e^{\tau\D} \Gamma (u)  = \Gamma (e^{\tau\D} u)\ ,\quad
\D = x \frac{\partial}{\partial x}\ ,\quad
e^{\tau\D} u(x) = u(xe^\tau),\quad \tau\in\real\ .
\end{equation}
\end{itemize}
}
}

No specific information about $\Gamma$ beyond these three conditions
will be used in sections 6 and 7.

It was already shown in Theorem~\ref{thm5.2} that the conditions
{\bf a)}  and {\bf b)} guarantee existence and uniqueness of the solution
$u(x,t)$ to the initial value  problem~\eqref{eq4.6}-\eqref{eq4.7}.
The property {\bf b)} yields the estimate \eqref{eq5.11} that is very important
for large time asymptotics, as we shall see below. The property   {\bf c)}
suggests a special class of self-similar solutions to Eq.~\eqref{eq4.6}.

Next, we recall  the usual  meaning of the notation $y = O(x^p)$
(often used below): $y = O(x^p)$
 if and only if  there exists a  positive constant
$C$ such that
\begin{equation}\label{big-o}
 |y(x)| \le C x^p \ \ \ \   \text{for \ any} \ \ \
x\ge 0.
\end{equation}

In order to study long time stability properties to solutions whose
 initial data differs in terms of $ O(x^p) $, we will need some
 spectral properties of  the linear operator $L$.

\begin{definition}\label{spectra}
Let $L$ be the positive  linear operator
 given in Eqs.~\eqref{eq5.2}, \eqref{eq5.3},then
\begin{equation}\label{eq6.6}
L x^p = \lambda (p) x^p\ ,\qquad \ \ \ \
0< \lambda (p) = \int_0^\infty da\, K(a) a^p <\infty\ , \quad\ p\geq 0\, ,
\end{equation}
and the spectral function $\mu(p)$  is defined by
\begin{equation}\label{spec-func}
\mu(p) = \frac{\lambda (p)-1} p\ .
\end{equation}
\end{definition}

An immediate consequence of properties  {\bf a)} and
{\bf b)}, as stated in \eqref{eq6.1}, is that
one can obtain a criterion for a  point-wise in $x$
estimate  of   the difference of two solutions to
the initial value problem~\eqref{eq4.6}  yielding
 decay properties depending on the spectrum of $L$.

 \begin{lem}\label{lem6.2n}
 Let $u_{1,2}(x,t)$ be any  two classical solutions of the problem
 \eqref{eq4.6} with  initial data   satisfying the conditions
 \begin{equation}\label{eq6.7n}
|u_{1,2}(x,0)| \leq1, \ \ \ \  \ |u_1(x,0) - u_2(x,0)| \leq C\,x^p\, ,\ \ x\geq0
 \end{equation}
 for some positive constant $C$ and   $p$.
 Then
  \begin{equation}\label{eq6.8n}
 |u_1(x,t) - u_2 (x,t)| \le C{x^p}  \,  e^{-t(1-\lambda(p))} \, , \qquad\text{for\  all}\ \ t\geq 0
\end{equation}
 \end{lem}

 \begin{proof}
The existence and uniqueness of $u_{1,2}(x,t)$ follow from
Theorem~\ref{thm5.2}. Estimate~ \eqref{eq5.11} (a consequence
 from the $L$-Lipschitz condition!) yields
 \begin{equation}\label{eq6.5}
 |u_1(x,t) - u_2 (x)|
 \le e^{-t} e^{Lt} w(x)\, ,\ \ \text{with}\ \ w(x)= |u_1(x,0) - u_2(x,0)| \ .
 \end{equation}

The operator $L$ from ~\eqref{eq5.2} is positive, and therefore monotone.
Hence we obtain
\begin{equation*}\label{eq6.6.bis}
e^{tL}w(x) = \sum_n \frac{t^n}{n!}\, L^n\,w(x) \leq
C\,e^{tL} \, x^p = C e^{\lambda(p)\,t}x^p,
\end{equation*}
that completes the proof.
 \end{proof}

\begin{corollary}\label{cor00}
The minimal constant $C$ for which condition \eqref{eq6.7n}
is satisfied is
\begin{equation}\label{eq6.9n}
C_0= \sup_{x\geq 0} \frac{|u_1(x,0) - u_2(x,0)| }{x^p}=
\left\|\frac{u_1(x,0) - u_2(x,0)| }{x^p} \right\|\, ,
\end{equation}
and the following estimate holds
\begin{equation}\label{eq6.10n}
\left\|\frac{u_1(x,t) - u_2(x,t)| }{x^p} \right\| \leq
e^{-t(1-\lambda(p))}\left\|\frac{u_1(x,0) - u_2(x,0)| }{x^p} \right\|
\end{equation}
for any $p>0$.
\end{corollary}
\begin{proof}
It follows directly from Lemma~\ref{lem6.2n}.
\end{proof}

We note that the result similar to Lemma~\ref{lem6.2n} was first
obtained in \cite{BC-03} for the inelastic Boltzmann equation whose
Fourier transform  is given in example {\bf (C)},
Eq.~\eqref{eq2.10}. Its corollary in the form similar to
\eqref{eq6.10n} for equation ~\eqref{eq2.10} was stated later in
\cite{BCT-03} and was interpreted there as ``the contraction
property of the Boltzmann operator'' (note that the left hand side
of Eq.\eqref{eq6.10n} can be understood as a {non-expansive
distance} any between two solutions).

However, independently of the terminology, the key reason for
estimates \eqref{eq6.8n}-\eqref{eq6.10n} is the L-Lipschitz property
of the operator $\Gamma$ as defined in \eqref{eq6.2}. It is actually
remarkable that the large time asymptotics of $u(x,t)$, satisfying
the problem \eqref{eq4.6} with such $\Gamma$, can be explicitly
expressed through spectral characteristics of the linear operator
$L$.

Hence, in order to study the large time asymptotics of   $u(x,t)$ in
more detail, we distinguish two different kinds of asymptotic
behavior:
\begin{itemize}
\item[{\bf 1)}] convergence to stationary solutions,
\item[{\bf 2)}] convergence to self-similar solutions provided the
  condition ({\bf c}), of the main properties on $\Gamma$, is satisfied.
\end{itemize}

The case {\bf 1)} is relatively simple. Any stationary solution
$\bar u(x)$ of the problem  \eqref{eq4.6} satisfies the equation
 \begin{equation}\label{eq6.11n}
\Gamma(\bar u)=\bar u\, , \qquad  \bar u \in C(\real_+)\, , \ \ \|\bar u\|\leq 1\, .
 \end{equation}

If the stationary solution $\bar u(x)$  does exists (note, for example, that
$\Gamma(0)=0$ and $\Gamma(1)=1$ for $\Gamma$ given in Eqs.~\eqref{eq4.3})
then the large time asymptotics of some classes of initial  data $u_{0}(x)$ in
\eqref{eq4.6} can be studied directly on the basis of Lemma~\ref{lem6.2n}.
 It is enough to assume that ${|u_0(x) - \bar u(x)| }$ satisfies \eqref{eq6.7n}
 with $p$ such that $\lambda(p) <1$. Then $u(x,t) \to \bar u(x)$ as
 $t\to\infty$, for any $x\geq0$.

This simple consideration, however, does not answer at least two questions:
\begin{itemize}
\item[{\bf A)}] What happens with $u(x,t)$ if the inequality~\eqref{eq6.7n}
 for  ${|u_0(x) - \bar u(x)| }$ is satisfied with such $p$  that $\lambda(p)>1$?
\item[{\bf B)}] What happens with $u(x,t)$ for large $x$
 (note that the estimate \eqref{eq6.8n} becomes trivial if $x\to\infty$).
\end{itemize}

In order to address these questions we consider a special class of solutions
 of Eq.~\eqref{eq4.6}, the so-called self-similar solutions. Indeed the
 property {\bf c)} of $\Gamma$ shows that  Eq.~\eqref{eq4.6} admits a
class of formal solutions $u_s(x,t)=w(x\, e^{{\mu_*}t}) $ with
some real $\mu_*$.
It is convenient for our goals to use a terminology that slightly differs from
 the usual one.

\begin{definition}\label{self-similar} The function
 $w(x)$ is called a  self-similar solution associated with the initial
 value problem~\eqref{eq4.6}
if it satisfies the problem
\begin{equation}\label{eq6.8}
\mu_* \D w + w = \Gamma (w)\ ,\qquad \|w\| \le 1\ ,
\end{equation}
in the notation of Eqs.~\eqref{eq6.2}, \eqref{eq4.3}.
\end{definition}

Note that the convergence of solutions $u(x,t)$ of the initial value
problem~\eqref{eq4.6}
to a stationary solution $\bar u(x)$
can be considered as a special case of the self-similar asymptotics
with $\mu_*=0$.

 Under the assumption that self-similar solutions exists
(the existence is proved in the next section), we prove the fundamental
result on the convergence  of solutions $u(x,t)$ of the initial value
problem~\eqref{eq4.6}  to self-similar ones (sometimes called in the
 literature {\sl  self-similar stability}).

\begin{lem}\label{lem6.1}
We assume that
\begin{itemize}
\item[{\bf i)}] for some $\mu_*\in\real$, there exists a classical
(continuously
differentiable if $\mu_*\ne0$) solution $w(x)$ of Eq.~\eqref{eq6.8} such
that $\|w\|\le 1$;
\item[{\bf ii) }] the initial data $u(x,0)= u_0$ in the problem \eqref{eq4.6}
 satisfies
\begin{equation}\label{eq6.9}
u_0 = w + O(x^p)\ ,\qquad \|u_0\| \le 1\ ,\text{ for } \ p>0\ \text{ such that }
\mu (p) <\mu_* ,
\end{equation}
where $\mu(p)$  defined in
 \eqref{spec-func} is  the spectral function
 associated to the operator $L$.
\end{itemize}
Then
\begin{equation}\label{eq6.12.1}
|u(xe^{-\mu_* t},t) - w(x)| = O({x^p}) e^{-pt(\mu_* -\mu(p))} \,
\end{equation}
 and therefore
\begin{equation}\label{eq6.12}
\lim_{t\to\infty} u(xe^{-\mu t}, t) = w(x)\ ,\qquad x\ge 0\ .
\end{equation}
\end{lem}

\begin{proof}
By assumption, the function  $u_2(x,t)=w(x\, e^{{\mu_*}t}) $
satisfies Eq.~\eqref{eq4.6}. Let  $u_1(x,t)$ be a solution of the problem
\eqref{eq4.6} such that $u_1(x,0)=u_0(x)$. Then, by Lemma~\ref{lem6.2n} and
by assumption {\bf ii)} we obtain
 $$|u_1(x,t) - w(xe^{{\mu_*} t})| = C\, x^p  e^{-(1-\lambda(p))\, t}\ ,$$
for some constant $C>0$ and all $x\geq 0\, , t\geq 0 \, .$

We can change in this inequality $x$ to $\tilde x\, e^{-{\mu_*} t}$,
then
$$
|u_1(x  e^{-{\mu_*} t },t) - w(x)| = C\, x^p  e^{-(p\mu_* + 1-\lambda(p))\, t}\ ,
$$

where the tildes are omitted. Note that $u_1(x,t)=u(x,t)$
in the formulation of the lemma and that
 $p\mu_* + 1-\lambda(p)= p(\mu_*-\mu(p)) $ in the notation \eqref{spec-func}.

Hence, the estimate \eqref{eq6.12.1} is proved. Eq.~\eqref{eq6.12} follows from
\eqref{eq6.12.1} since $\mu_* < \mu(p)$.
So the proof is completed.
\end{proof}

\begin{remark} Lemma~\ref{lem6.1}
 shows how to find a domain of attraction of any
self-similar solution provided the self-similar
 solution is itself  known.
It is remarkable that the domain of attraction can be expressed
in terms of just  the {\sl spectral function}  $\mu(p),\ p>0$,
defined in~\eqref{spec-func}, associated with
the linear operator $L$ for which the operator $\Gamma$ satisfies
the $L$-Lipschitz condition.

Generally speaking, the equality  \eqref{eq6.12} can be also
fulfilled for some other values of $p$ with $\mu (p) >\mu_*$
 in Eq.~\eqref{eq6.9},
but, at least, it always holds if $\mu (p)<\mu_*$.
\end{remark}

\medskip

We shall need some  properties of the
{\sl spectral function}  $\mu (p)$.
Having in mind further applications, we formulate these properties in terms
of the operator $\Gamma$ given in Eqs.~\eqref{eq4.3}, though they depend only
 on $K(a)$ in Eqs.~\eqref{eq6.6}.

\begin{lem}\label{lem6.2}
The spectral function $\mu (p)$ has the following properties:
\begin{itemize}
\item[{\bf i)} ] It is positive and unbounded as $p\to 0^+$, with
 asymptotic behavior given by
\begin{equation}\label{eq6.13}
\mu (p) \approx \frac{\lambda (0) -1}p\ ,\qquad
p\to 0\ ,
\end{equation}
where, for $\Gamma$ from \eqref{eq4.3}
\begin{equation}\label{eq6.14}
\lambda(0) = \int_0^\infty da\, K (a)
= \sum_{n=1}^N \alpha_n n\ge 1\ ,\qquad
\sum_{n=1}^N \alpha_n =1\ ,\ \alpha_n \ge 0\ ,
\end{equation}
and therefore $\lambda (0)=1$ if and only if the operator  $\Gamma$
\eqref{eq4.3} is linear $(N=1)$;
\item[{\bf ii)}] In the case of a multi-linear $\Gamma$ operator,
 there is not more than one point $0<p_0 <\infty$, where
the spectral function $\mu(p)$ achieves its minimum, that is,
$\mu'(p_0)=\frac{d\,\mu}{d\,p} (p_0) =0$, with  $\mu (p_0) \le \mu (p)$ for any $p>0$,
provided  $N\ge 2$ and $\alpha_N >0$.
\end{itemize}
\end{lem}

\begin{proof}
Eqs. ~\eqref{eq6.13}, \eqref{eq6.14}
follow directly from the definition
\eqref{spec-func} of $\mu (p)$ and from Eqs.~\eqref{eq5.3}, \eqref{eq6.6}.

The statement {\bf ii)} follows, first, from the
convexity of $\lambda (p)$  since
\begin{equation}\label{eq6.15}
\lambda'' (p) = \int_0^\infty da\, K(a) a^p(\ln a)^2 \geq 0 \ ,
\end{equation}
and from the identity
\begin{equation}\label{eq6.15.1}
\mu'(p) = \frac{\psi (p)}{p^2}\ ,\qquad
\psi (p) = p\lambda' (p) - \lambda (p) +1\ .
\end{equation}
We note that $\psi (p)$ in~\eqref{eq6.15.1} is a monotone increasing function
of $p$,$(\psi' = p\lambda'' \ge0)$ and therefore it has not more than one
zero, at say,  $p=p_0 >0$.
Now, if $N\geq 2, \alpha_N>0 $ (i.e. $\Gamma$ is non-linear),
 then $p = p_0$ is also
a  minimum point for $\mu (p)$ since from Eq.~\eqref{eq6.13}
$\mu(p)\to +\infty$ as $p\to 0$ and thus
$\mu' (p) <0$ for $p\to0$.
This completes the proof.
\end{proof}

\begin{remark} From now on, we
shall always assume below that the operator $\Gamma$ from~\eqref{eq4.3}
is multi-linear.
Otherwise it is easy to see that the problem \eqref{eq6.8} has no
solutions (the condition $\|w\| \le 1$ is important!) except for  the trivial
ones $w=0,1$.
\end{remark}

The following corollaries are readily obtained from
lemma~\ref{lem6.2}.

\begin{corollary}\label{cor0}
For the case of a non-linear $\Gamma $ operator, i.e. $N\geq 2$,
the spectral function $\mu(p)$ is always monotone decreasing in the interval
$(0,p_0)$, and $\mu(p) \geq \mu(p_0)$ for $0 <p<p_0$. This implies that
 there exists
a unique inverse function
 $\tp(\mu):(\mu(p_0), +\infty) \to (0,p_0) $, monotone decreasing
in its domain of definition.
\end{corollary}
\begin{proof} It follows immediately from Lemma~\ref{lem6.2}, part
{\bf ii)} and its proof.
\end{proof}

\begin{corollary}\label{cor1}
There are precisely four different kinds of qualitative behavior of
$\mu (p)$ shown on Fig.1.
\end{corollary}

\begin{proof}
There are two options: $\mu (p)$ is a monotone decreasing function
(Fig.1~(a)) or $\mu (p)$ has a minimum at $p=p_0$ (Fig.1~{(b,c,d)}).
In  case  Fig.1~{(a)} $\mu (p) >0$ for all $p>0$ since $\mu (p) >
1/p$. The asymptotics of $\lambda (p)$ \eqref{eq6.6} is clear:
\begin{align} % requires amsmath; align* for no eq. number
   \text{{\bf (1)}}\qquad &\lambda (p) \xrightarrow[p\to\infty]{}
\lambda_\infty
   \in \real_+\ \text{ if }\  \int_{1^+}^\infty da\, K(a) = 0\ ;
\label{eq6.16} \\
   \text{{\bf (2)}}\qquad &\lambda (p) \xrightarrow[p\to\infty]{}\
\infty \text{ if }\
   \int_{1^+}^\infty da\, K(a) >0\ .
\end{align}

In the case {\bf (1)} when $\mu (p) \to\infty$ as $p\to 0$, two possible
pictures (with and without minimum) are shown on Fig.1~{(b)} and
Fig.1~{(a)} respectively.
In case {\bf (2)}, from Eq.~\eqref{eq6.6} it is clear that
$\lambda (p)$ grows exponentially for large $p$,
therefore $\mu (p) \to \infty$ as $p\to \infty$.
Then the minimum always exists and we can distinguish two cases:
$\mu (p_0) <0$ (Fig.1~{(d}) and $\mu (p_0) >0$ (Fig.1~{(c)})
\end{proof}
\begin{figure}
 \centering
\label{figure1}
\includegraphics[width=7cm]{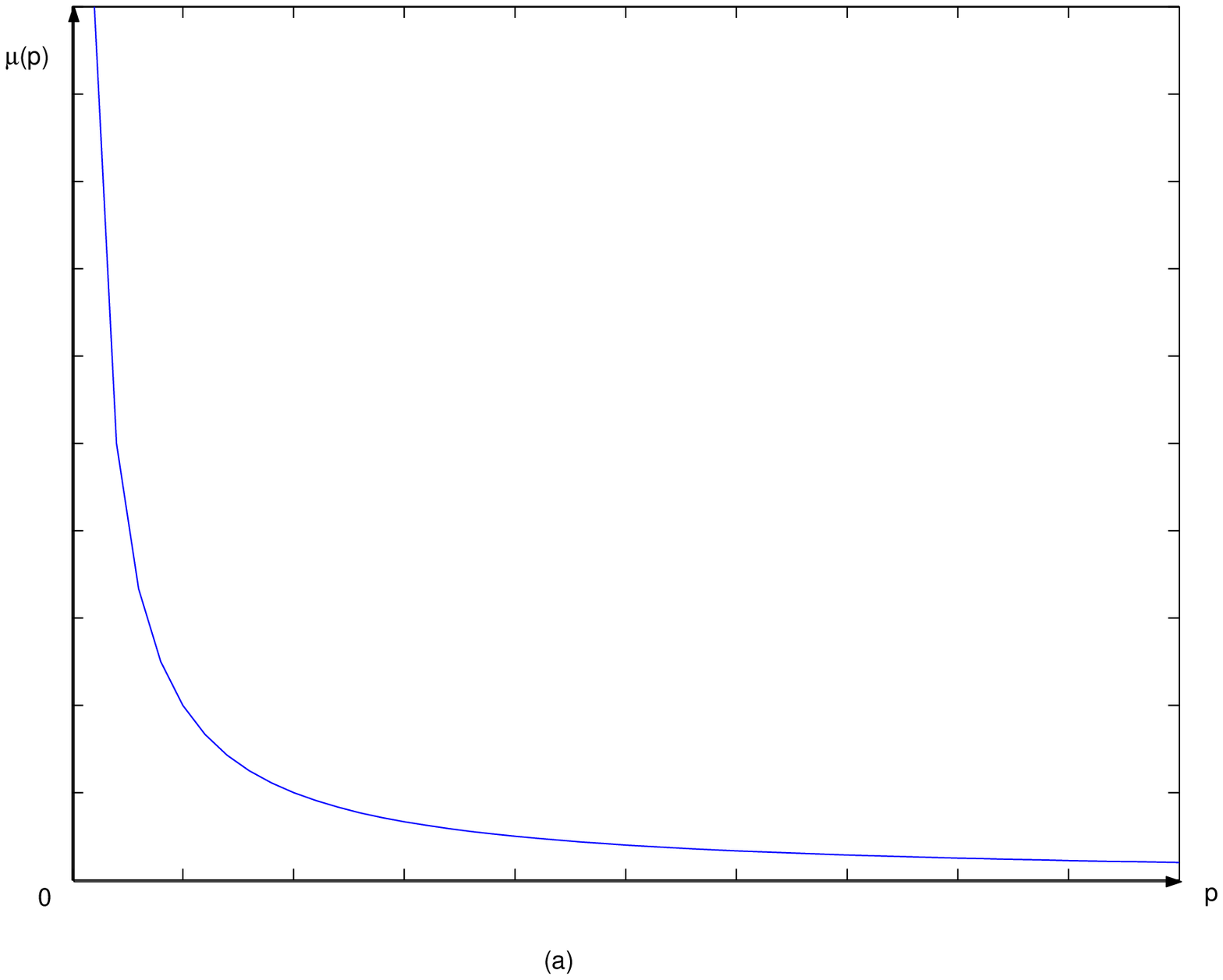}\qquad
\includegraphics[width=7cm]{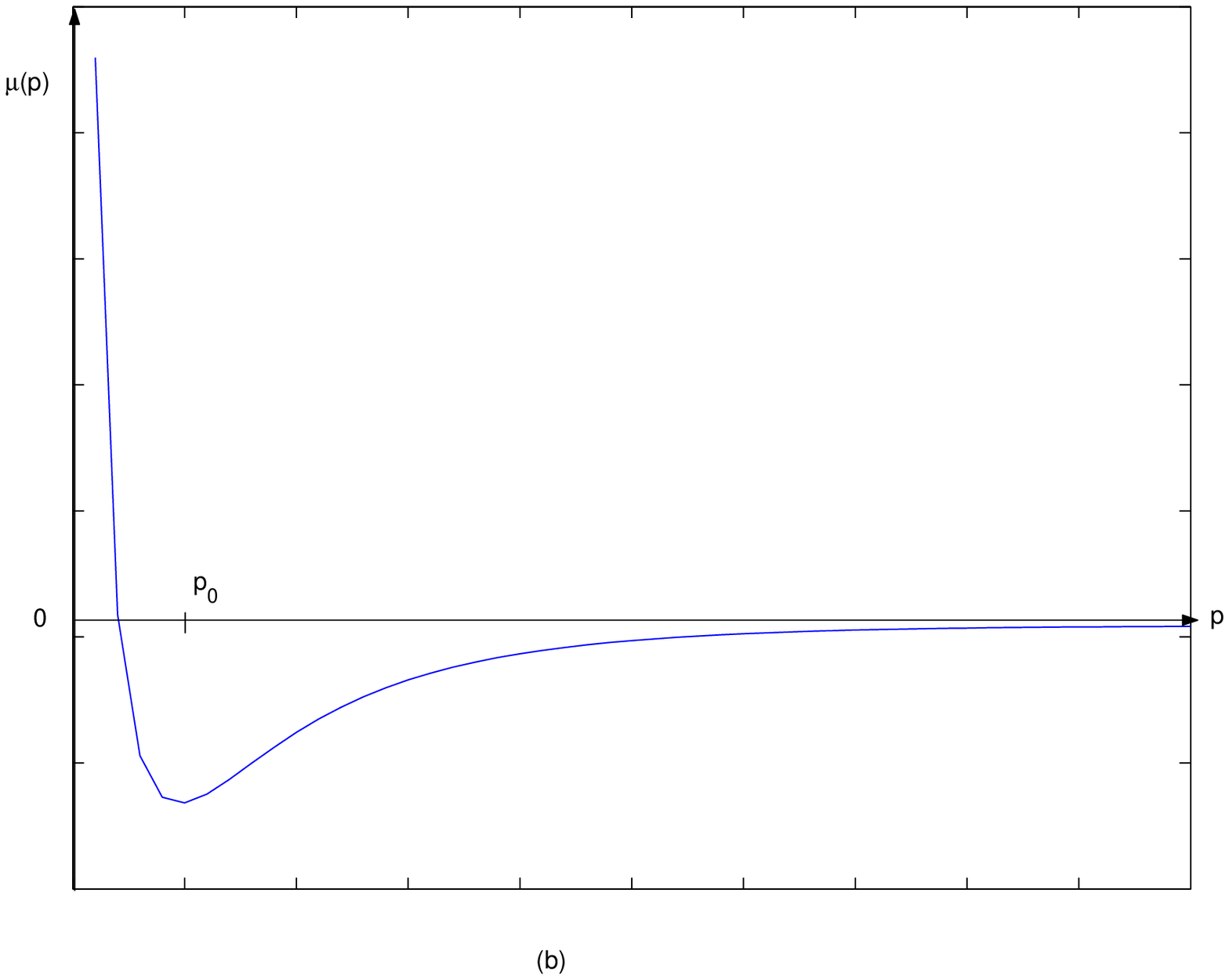}\\
\ \\
\includegraphics[width=7cm]{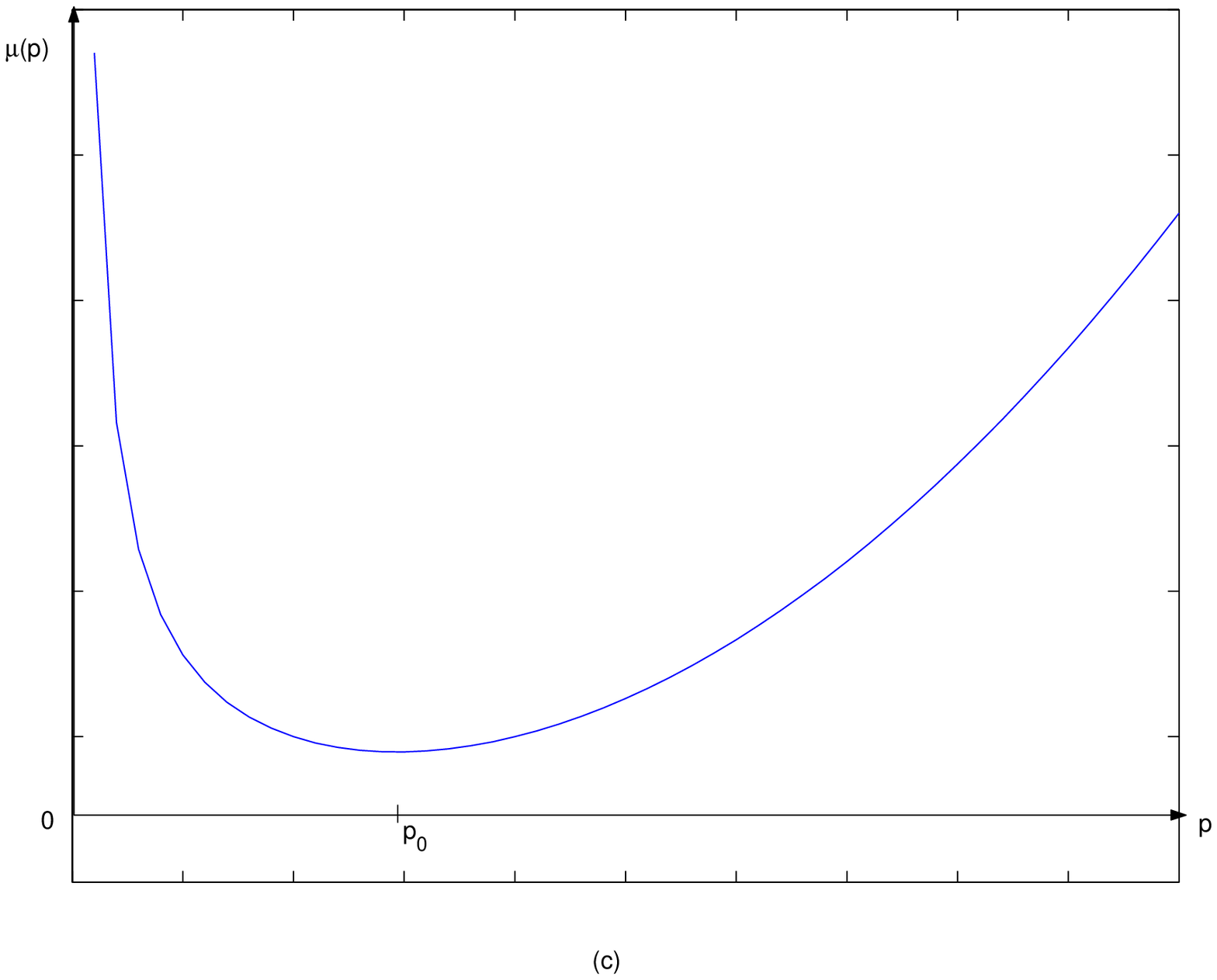}\qquad
\includegraphics[width=7cm]{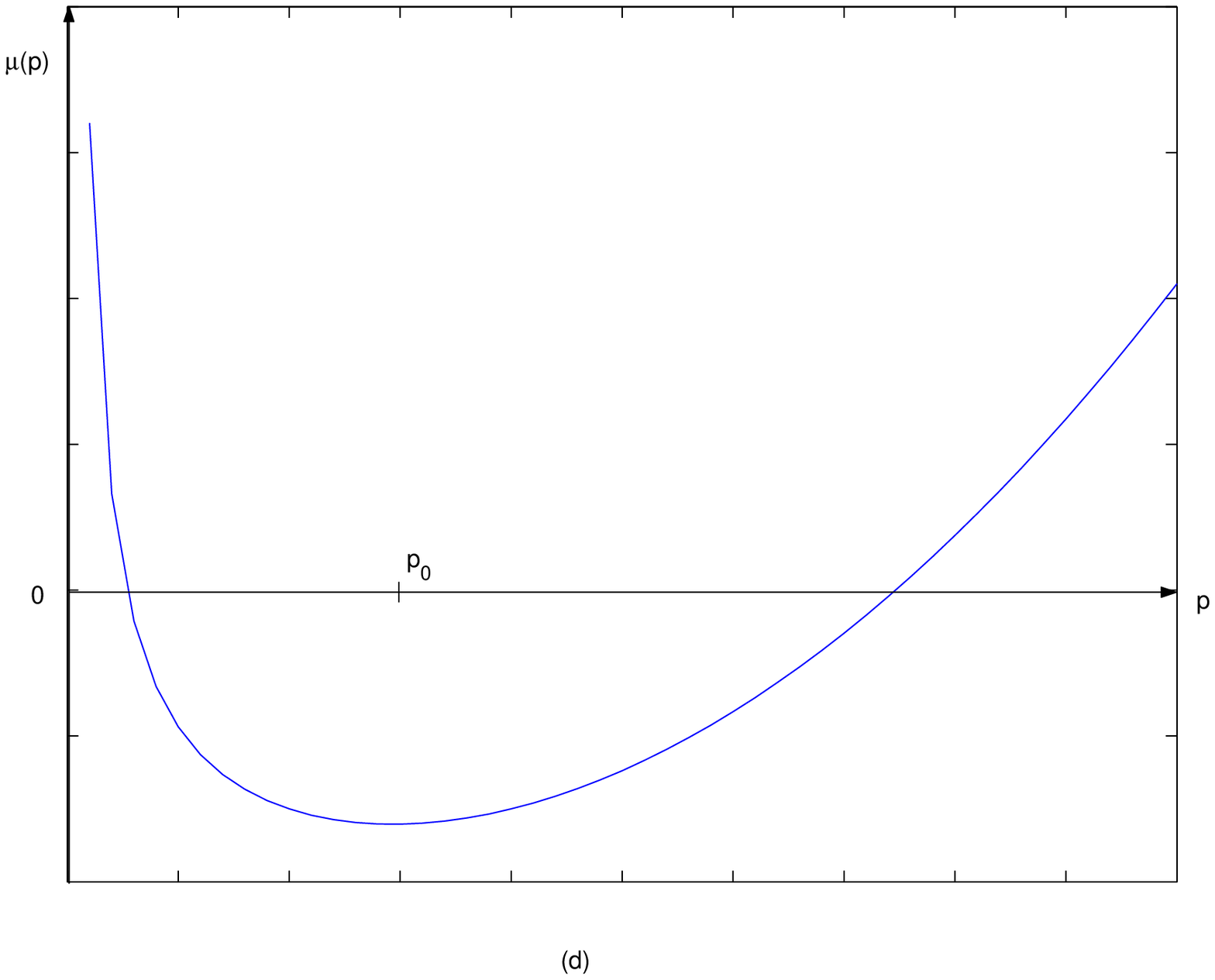}
\caption{Possible profiles of the spectral function $\mu(p)$
}
\end{figure}

We note that, for Maxwell models {\bf (A), (B), (C)} of Boltzmann equation
(Sections~2 and 3), only cases {(a)} and {(b)}
of Fig.1 can be possible
(actually this is the case {(b)}) since the condition \eqref{eq6.16} holds.
Fig.1 gives a clear graphic representation of the domains of
attraction of self-similar solutions (Lemma~\ref{lem6.1}):
it is sufficient to draw the line $\mu (p) = \mu_* = {constant}$, and
to consider a $p$ such  that the graph of $\mu (p)$ lies below this line.

Therefore, the following corollary follows directly from the properties
of the spectral function $\mu(p)$, as characterized by the  behaviors in
 Fig.1, where
we assume that
 $\mu (p_0)=0$ for $p_0= \infty$, for the case shown on Fig.1~{(a)}.

\begin{corollary}\label{cor2}
Any self-similar solution $u_s(x,t) = w(xe^{\mu_* t})$ with $\mu (p_0)
<\mu_* <\infty$ has a non-empty domain of attraction, where $p_0$ is the unique
(minimum) critical point of the spectral function $\mu(p)$.
\end{corollary}

\begin{proof}
We use Lemma~\ref{lem6.1} part {\bf ii)} on any initial state
$u_0= w + O(x^p)$ with $p>0$ such that $\mu (p_0) \leq \mu (p)
<\mu_* $. In particular, Eqs.~\eqref{eq6.12.1} and   \eqref{eq6.12}
show that the domain of attraction of  $ w(xe^{\mu_* t})$ contains any
 solution to the initial value problem~\eqref{eq4.6} with the initial
state as above.
\end{proof}

\medskip

The inequalities of the kind $u_1 - u_2 = O (x^p)$ for\ any $p>0 $
such  that $\mu (p) <\mu_*$,   for any fixed $\mu_* \geq \mu(p_0)$
play an important role for the self-similar stability. We can use
specific properties of $\mu (p)$ in order to express such
inequalities in a more convenient form.

\begin{lem}\label{lem6.3}
 For any given $\mu_* \in (\mu (p_0),\infty)$ and
$u_{1,2}(x)$ such that $\|u_{1,2}\| <\infty$,
 the following two statements are equivalent:
\begin{itemize}
\item[{\bf i)}]
there exists  $p>0$ such that
\begin{equation}\label{eq6.18}
u_1 - u_2 = O (x^p)\ ,\qquad  \text{with}\ \
\mu (p) <\mu_*\, .
\end{equation}

\item[{\bf ii)}] There exists $\ep >0$ such that
\begin{equation}\label{eq6.19}
u_1 - u_2 = O(x^{{\tp}(\mu_*) +\ep})\ ,
\quad \ \text{ with}\ \
 \tp(\mu_*) < p_0\ ,
\end{equation}
where $\tp(\mu)$ is the inverse to $\mu (p)$ function, as defined
in {\bf Corollary~\ref{cor0}}.
\end{itemize}
\end{lem}

\begin{proof}
Let property {\bf i)} holds,  then  recall from {\bf Corollary~\ref{cor0} }
that $\mu (p)$ is monotone on the interval $0<p\le p_0$,
so  its  inverse function $\tp(\mu)$ is defined uniquely.

It is clear, as it can be seen in  from Fig.1,
 that the condition $\mu (p)<\mu_*$ are satisfied
only for some $p>\tp(\mu_*)$, therefore  inequality \eqref{eq6.19}
with $\ep = p-\tp(\mu_*)$ follows directly from \eqref{eq6.18}.

Conversely, if  {\bf ii)} holds, first note that for
 any pair of uniformly bounded functions (note that
$\|u_{1,2}\| <\infty$ by assumption) which
satisfy  inequality
$$|u_1(x) - u_2(x)| < C\, x^q\ ,\qquad C = {const.}\ ,$$
 for some $q>0$,
then the same inequality holds with any  $p$ such that
 $0<p<q$ and perhaps
another constant.
Therefore, if the condition \eqref{eq6.19} is satisfied,
 then one  can
always find a  sufficiently small $0<\ep_1 \le \ep$ such that
taking $p= \tp (\mu_*) +\ep_1 $
condition \eqref{eq6.18}  is fulfilled.
This completes the proof.
\end{proof}

\medskip

Finally, to conclude this section,
 we show  a general property of the initial value  problem~\eqref{eq4.6}
for any non-linear $\Gamma$ operator
 satisfying conditions {\bf a)} and {\bf b)}
given in Eqs.~\eqref{eq6.1bis} and \eqref{eq6.1} respectively.
This property gives the control to the point-wise difference
of any  two rescaled
 solutions to \eqref{eq4.6} in the unit sphere of B , whose
initial states differ by $O(x^p)$. It is formulated as follows.

\begin{lem}\label{lem6.4}
Consider the problem \eqref{eq4.6}, where $\Gamma$ satisfies the
conditions {\bf (a)} and {\bf (b)}.
Let $u_{1,2} (x,t)$ are two solutions satisfying the initial
conditions $u_{1,2} (x,0) = u_0^{1,2} (x)$ such that
\begin{equation}\label{eq6.20}
\|u_0^{1,2}\| \le 1\ ,\quad
u_0^{1} - u_0^{2} = O(x^p)\ ,\quad
p>0\ .
\end{equation}
then, for any  real $\mu_*$,
\begin{equation}\label{eq6.21}
\Delta_{\mu_{*}} (x,t) = u_{1} (xe^{-\mu_* t},t) - u_{2} (xe^{-\mu_* t},t)
= O( x^p) e^{-pt[\mu_* -\mu (p)]}
\end{equation}
and therefore
\begin{equation}\label{eq6.22}
\lim_{t\to \infty} \Delta_{\mu_{*}} (x,t) = 0\ ,\qquad x\ge0\ ,
\end{equation}
for any $\mu_* >\mu (p)$.
\end{lem}

\begin{proof}
The proof is a repetition of arguments that led to
Eq.~\eqref{eq6.12.1}). In particular,
we obtain from Eqs.~\eqref{eq5.11}, \eqref{eq6.8n} the estimate
$$|u_1 (x,t) - u_2 (x,t)| = x^p\, e^{t[\lambda (p)-1]} \,
\|\frac{u_0^{1} - u_0^{2}}{x^p}\|  $$
and then change $x$ to $xe^{-\mu_* t}$.
This leads to Eq.~\eqref{eq6.21} in the notation \eqref{spec-func} and
this completes the proof.
\end{proof}

\begin{remark} There  is an important point to understand here:
Lemmas~\ref{lem6.2n} and~\ref{lem6.1}  hold for any operator
$\Gamma$ that satisfies just the  two properties {\bf (a) }and
{\bf(b)} stated in \eqref{eq6.1bis} and \eqref{eq6.1}. It says that,
in some sense, a distance between any two solutions with initial
conditions satisfying Eqs.~\eqref{eq6.20} tends to zero as $t\to
\infty$, i.e. {\sl non-expansive distance}. Such terminology and
corresponding distances were introduced in \cite{GTW-95} for the
elastic Maxwell-Boltzmann with  finite initial energy, and  for
specific forms of Maxwell-Boltzmann models in \cite{BCT-06,PT05}. It
should be pointed out, however, that this {\sl contraction property}
may not say much about large time asymptotics of $u(x,t)$, unless
the corresponding self-similar solutions are known, for which the
operator $\Gamma$ must be invariant under dilations (so it satisfies
also
 property {\bf (c)} as well, as stated in \eqref{eq6.2}).
In such case one can use estimate~\eqref{eq6.22} to deduce the
pointwise convergence of characteristic functions $u(x,t)$ (Fourier
or Laplace transforms of positive distributions, or equivalently,
probability measures) in the form \eqref{eq6.12.1}, \eqref{eq6.12}
and the corresponding weak convergence in the space of probability
measures \cite{Fe}. We shall discuss these issues, in Sections~10
~and~11 below, in more detail.

\end{remark}

Therefore one must study  the problem of existence of self-similar
 solutions, which is  considered in the next
Section.

\vskip.5in

%%%%%%%%%%%%%%%%%%%%%%%%%%%%%%%%%%%%
%%%%%%%%  Section 7  %%%%%%%%%%%%%%%%%
%%%%%%%%%%%%%%%
\section{Existence of self-similar solutions}

The goal of this Section is to develop a criterion for existence,
uniqueness and self-similar asymptotics to the    problem
Eq.~\eqref{eq6.8} for any operator $\Gamma$  that satisfies
conditions  {\bf (a)}, {\bf (b)} and {\bf (c)} from Section~6, with
the
 corresponding spectral function $\mu(p)$ defined in
\eqref{spec-func}.

Theorem~\ref{thm7.1} below shows the criterion for
 existence and uniqueness of
self-similar solutions for any operator $\Gamma$ that satisfies just
conditions  {\bf (a)} and {\bf (b)}.  Then  Theorem~\ref{thm7.2}
follows,  showing a general criteria to self-similar asymptotics for
the problem \eqref{eq4.6} for any operator $\Gamma$  that satisfying
conditions  {\bf (a)}, {\bf (b) } and {\bf (c)} and that $p_0>1$,
for $\mu(p_0)=\min_{P>0} \mu(p)$, in order to study the initial
value problem with finite energy.

\medskip

We consider Eq.~\eqref{eq6.8} written in the form
\begin{equation}\label{eq7.1}
\mu_* xw' (x) + w(x) = g(x)\ ,\qquad g = \Gamma (w)\ ,\ \mu_* \in
\real\ ,
\end{equation}
and, assuming that $\|w\| <\infty$, transform this equation to the
integral form.
It is easy to verify that the resulting integral equation reads
\begin{equation}\label{eq7.2}
w(x) = \int_0^1 d\tau\, g(x\tau^{\mu_*})
\end{equation}

We prove the following result.
\begin{thm}\label{thm7.1}
Consider Eq.~\eqref{eq7.1} with arbitrary $\mu_*\in\real$ and the
operator $\Gamma$ satisfying the conditions {\bf (a)} and {\bf (b)}
from Section~6. Assume that there exists a continuous function
$w_0(x)$, $x\ge0$, such that
\begin{itemize}
\item[{ i)}] \ \ \ \ \ \ \ \ \ $\|w_0\| \le 1$ and

\item[{ ii)}]
%\vskip-36pt
\begin{equation}\label{eq7.10}
 \int_0^1 dt\, g_0 (xt^\mustar) = w_0 (x) + O(x^p)\ ,\qquad \qquad
g_0 = \Gamma (w_0)\ ,\hfill
\end{equation}
\newline
with some $p>0$ and $p\,\mu_* > -1$  satisfying the inequality
\begin{equation}\label{eq7.11}
\mu (p) = \frac1p \Big[ \int_0^\infty da\, K(a) a^p -1\Big] <\mu_*\
.
\end{equation}
\end{itemize}
\begin{itemize}
\item[\ ] Then, there exists a classical solution $w(x)$
of Eq.~\eqref{eq7.1}.
\item[\ ] The solution is unique in the class of
continuous functions satisfying conditions
\begin{equation}\label{eq7.12}
\|w\| \le 1\ ,\qquad w(x) = w_0 (x) + O(x^{p_1})\ ,
\end{equation}
with any $p_1$ such that $\mu (p_1) <\mu_*$.
\end{itemize}
\end{thm}

 \begin{proof}
The existence is proven by the following iteration procedure. We
choose an initial approximation $w_0\in U$ such that $\|w_0\| \le 1$
and consider the iteration scheme
\begin{equation}\label{eq7.3}
w_{n+1} (x) = \int_0^1 d\tau\, g_n(x\tau^\mustar) \ ,\qquad\qquad
g_n = \Gamma (w_n)\ ,\ n= 0,1,\ldots\ .
\end{equation}

Then, property {\bf (a)} of $\Gamma$, $\|w_n\| \le 1$ for all $n\geq
1$ and
$$|w_{n+1} (x) - w_n(x)|
\le \int_0^1 d\tau\, |g_n(x\tau^{\mu_*}) - g_{n-1} (x\tau^{\mu_*})|\
,\qquad n\ge 1\, .$$

By using the inequality \eqref{eq6.1} (i.e. property {\bf (b)} of
$\Gamma$),  we control the right hand side of the above inequality
by, recalling the definition of the linear operator $L$
from~\eqref{eq5.2}
\begin{multline}\label{eq7.3.1}
|w_{n+1} (x) - w_n (x)| \le  \int_0^1 d\tau\,L(|w_n - w_{n-1}|)
(x\tau^\mustar) \\
= \int_0^1 d\tau\, \int_0^\infty da\, K(a) |w_n - w_{n-1}|
(ax\tau^\mustar) \ .
\end{multline}

Next,   by assumption ii), initially
\begin{equation}\label{eq7.4}
w_1 = w_0 + O(x^p)\ ,\qquad\mbox{with }\qquad  p\mustar >-1\ ,\ p>0\
,
\end{equation}
or, equivalently,
$$|w_1 (x) - w_0(x)| \le Cx^p\ ,\qquad x\ge 0\ ,\ C= {const.} \ ,$$
then, recalling the definition  for $\lambda (p)$ given in
\eqref{eq6.6}, we can control the right hand side of \eqref{eq7.3.1}
by
\begin{equation*}
x^p\, \int_0^\infty da\, K(a) a^p  \int_0^1 \tau^{p\mustar)} d\tau =
x^p \frac{\lambda (p)}{1+p \mustar}\ .
\end{equation*}
 Therefore,  we estimate the
left hand side of \eqref{eq7.3.1} by
\begin{equation}\label{eq7.5}
|w_{n+1} (x) - w_n(x)| \le C\gamma^n (p,\mustar) x^p\ ,\qquad \gamma
(p,\mustar) = \frac{\lambda (p)}{1+p \mustar}\ .
\end{equation}

Then,  from condition ii)  $\gamma (p,\mu_*)>0$ since $p\mu_* >-1$.
Also, from condition ii), \eqref{eq7.11}, $\mu(p) = \frac{\lambda
(p)-1}p<\mustar$  implies $0< \gamma(p,\mu_*) <1$.

Therefore, there exists a point-wise limit
\begin{equation}\label{eq7.6}
w(x) = \lim_{n\to\infty} w_n(x)
\end{equation}
satisfying the inequality
\begin{equation}\label{eq7.7}
|w(x) - w_0(x)| \le \sum_{n=0}^\infty |w_{n+1} - w_n(x)| \le
\frac{C}{1-\gamma (p,\mustar)}\ x^p\ .
\end{equation}

Estimate \eqref{eq7.5} with $\gamma <1$ shows that the
convergence  $w_n(x) \to w(x)$ is uniform on any interval
$0\le x\le R$, for any  $R>0$.
Therefore $w(x)$ is a continuous function, moreover $\|w\| \le 1$
since $\|w_n\| \le 1$ for all $n=0,1,\ldots$.

The next step is to prove that the limit function  $w(x)$ from
\eqref{eq7.6} satisfies Eqs.~\eqref{eq7.1}, or equivalently,
\eqref{eq7.2}. We note that
\begin{equation}\label{eq7.8}
\begin{split}
|g_{n+1} (x) - g_n(x)|
& \le \int_0^\infty da\, K(a) |w_{n+1} (ax) - w_n (ax)|\le \\
& \le C\, \lambda (p) \gamma^n (p,\mustar) x^p\ .
\end{split}
\end{equation}

Therefore $g_n(x) \to g(x)$, where $g(x)\in C(\real_+)$ and $\|g\|
\le 1$, since $\|g_n\|\le1$ for all $n$. In addition, from the
continuity of the operator $\Gamma (u)$ for $\|u\| \le 1$ follows
that  $g= \Gamma (w)$,  and the transition to the limit in the right
hand side of Eq.~\eqref{eq7.2} is justified since $\|g_n\| \le 1$.
Hence, $w(x)$ satisfies Eq.~\eqref{eq7.2}.

When $\mustar\ne 0$ one also  needs  to check that Eq.~\eqref{eq7.1}
is satisfied. We note that, for any continuous and bounded $w_0
(x)$, all functions $w_n(x)$, $n\ge 1$, are differentiable for $x>0$
and their derivatives $w'_n(x)$ satisfy the equations (see
Eqs.~\eqref{eq7.1}, \eqref{eq7.3})
\begin{equation*}
\mustar \, x\,  w'_n(x)  = w_n(x) + g_{n-1} (x)\ ,\qquad
n=1,2,\ldots\ .
\end{equation*}

Hence,
\begin{equation*}
\mustar\,  x \, (w'_{n+1} - w'_n) = (w_n- w_{n+1}) + (g_n -
g_{n-1})\ ,
\end{equation*}
and, by using inequalities \eqref{eq7.5}, \eqref{eq7.8}, we obtain
\begin{equation*}
|\mustar| \,  |w'_{n+1} - w'_n| \le C\, \gamma^{p-1} (p,\mustar)\,
(\gamma (p,\mustar) + \lambda (p)) \, x^p\ , \qquad n\ge 1\ .
\end{equation*}

Therefore the sequence of derivatives $\{ w'_n(x),\, n=1,\ldots\}$
converges uniformly on any interval $\ep \le x\le R$. Hence, the
limit function $w(x)$ from \eqref{eq7.6} is differentiable for
$x>0$,
\begin{equation}\label{eq7.9}
w'(x) = O(x^{p-1})\ ,\qquad p>0\ ,
\end{equation}
and the equality \eqref{eq7.1} is also satisfied for $\mustar \ne0$.

Finally we note that the condition of convergence $0<\gamma
(p,\mustar)<1$, which is equivalent to the condition $\mu (p)
<\mustar$, (see Eq.~\eqref{spec-func}) that has already appeared in
Lemma~\ref{lem6.1}.

 It remains to prove the statement concerning the uniqueness of
 the limit function  $w(x)$ from \eqref{eq7.6}.
 If there are two solutions $w^{_{1,2}}$ satisfying \eqref{eq7.1}, then the
 integral equation \eqref{eq7.2} yields
 \begin{equation}\label{eq7.13}
 |w^\wsuprao(x) - w^\wsuprat (x)| \le \int_0^1 d\tau\, |g^\wsuprao(x\tau^\mustar) -
 g^\wsuprat(x\tau^\mustar )|\ ,\qquad
 g^{_{1,2}} = \Gamma (w^{_{1,2}})\ .
 \end{equation}

 Since $\| w^{_{1,2}}\| \le 1$, we obtain
 \begin{equation*}
 |w^\wsuprao(x) - w^\wsuprat (x)| \le C\, x^q\ ,\qquad q = \min (p,p_1)\ ,
 \end{equation*}
 where obviously $\mu (q) <\mustar$.
 Then we again apply the inequality \eqref{eq6.1} to the integral in
 Eq.~\eqref{eq7.2} and get the new estimate
 \begin{equation*}
 |w^\wsuprao (x) - w^\wsuprat(x)| \le C_1 x^q\ ,\qquad C_1 = \gamma (q,\mustar) C < C\ .
 \end{equation*}

 By repeating the same considerations as in the existence argument,  it follows that
 \begin{equation*}
 |w^\wsuprao (x) - w^\wsuprat (x)| \le C\gamma^n (q,\mustar) x^q\ ,\qquad
 \text{ with }\ \ \gamma (q,\mustar) < 1\ ,
 \end{equation*}
 for any integer $n\geq 0$.
 Therefore $w^\wsuprao(x) \equiv w^\wsuprat (x)$ and the proof is complete.
 \end{proof}

 Now we can combine Lemma~\ref{lem5.1}
 with Lemma~\ref{lem6.1}
 and prove the general statement related to the self-similar asymptotics
 for the problem \eqref{eq4.6}.

 \begin{thm}\label{thm7.2}
 Let $u(x,t)$ be a solution of the problem \eqref{eq4.6} with $\|u_0\|\le1$
 and $\Gamma$ satisfying the conditions {\bf (a)},
 {\bf (b)}, {\bf (c)} from Section~6.
% (\eqref{eq6.1}, \eqref{eq6.1bis}, \eqref{eq6.2}).
 Let $\mu (p)$ denote the spectral function \eqref{eq7.11} having its
 minimum (infimum) at $p = p_0$ (see Fig.1), the case $p_0 = \infty$
 is also included.
 We assume that there exists $p \in (0,p_0)$ and $0<\ep<p_0-p$ such that
 \begin{equation}\label{eq7.13a}
 \int_0^1 d\tau\, g_0 (x\tau^{\mu (p)})
 = u_0 (x) + O (x^{p+\ep})\ ,\qquad
 g_0 = \Gamma (u_0),\quad  \ep>0\ .
 \end{equation}
 Then
\begin{itemize}
\item[{ i)}] there exists a unique solution $w(x)$ of the
equation~\eqref{eq7.1} with $\mu_* = \mu (p)$  such that
\begin{equation}\label{eq7.14}
\|w\| \le 1\ ,\qquad w(x) = u_0 (x) + O(x^{p+\ep})\ ,
\end{equation}
\item[{ ii)}]
\begin{equation}\label{eq7.15}
\lim_{t\to\infty} u (x\,e^{-\mu (p)t} ,t) = w(x)\ ,\qquad x\ge 0\ ,
\end{equation}
where the convergence is uniform on any bounded interval
in $\real_+$ and
\begin{equation}\label{eq7.17}
u (x\, e^{-\mu (p)t} ,t) -  w(x)= O(x^{p+\ep}e^{-\beta(p,\ep)t} )\, ,
\end{equation}
with $\beta(p,\ep)= (p+\ep)(\, \mu(p) -\mu(p+\ep)\, ) >0$.
\end{itemize}
\end{thm}

\begin{proof}
If the condition \eqref{eq7.13a} is satisfied, we can take $w_0 = u_0(x)$
in the assumption { ii)}  of Theorem~\ref{thm7.1}.
Indeed, the function $\mu (p)$ is monotone decreasing for
$p\in (0,p_0)$ and therefore invertible.
We denote the inverse function by $p(\mu)$ and apply Lemma~\ref{lem6.3}
to the condition  {\bf ii)} of Theorem~\ref{thm7.1}.
Thus we obtain for any $\mu_* \in (\mu (p_0),\infty)$ and $w_0 = u_0$
\begin{equation}\label{eq7.16}
\int_0^1 d\tau\, g_0 (x \tau^{\mu_*}) = u_0 + O(x^{p(\mu_*) +\ep})\
,\qquad \ep>0\ .
\end{equation}
We note that the condition { ii)} cannot be fulfilled
for any $\mu_* \le\mu (p_0)$
since the set $\mu(p) <\mu_*$ is empty in such case.
Rewriting Eq.~\eqref{eq7.16} in the equivalent form with $\mu_* =\mu (p)$
we obtain the condition \eqref{eq7.14}.
Then the statement {\bf i)} follows from Theorem~\ref{thm7.1}.
Therefore we can apply Lemma~\ref{lem6.1} and obtain the limiting
equality \eqref{eq7.15} and the estimate \eqref{eq7.17}.
This completes the proof.
\end{proof}

\smallskip

Thus we obtain a general criterion \eqref{eq7.14} of the
self-similar asymptotics of $u(x,t)$ for a given initial condition
$u_0 (x)$. The criterion can be applied to the problem \eqref{eq4.6}
with any operator $\Gamma$ satisfying conditions {\bf (a)}, {\bf
(b)}, {\bf (c)} from Section~6. The specific class \eqref{eq4.3} of
operators $\Gamma$ is studied in Section~8. We shall see below that
the condition~\eqref{eq7.14} can be essentially simplified for such
operators.

\smallskip

\begin{remark}
One may expect a probabilistic connection between
Theorems~\ref{thm7.1} and \ref{thm7.2} on the relation of rates of
convergence between rates of relaxation to these selfsimilar states
and the convergence of the corresponding  Wild convolution sums
formulation for a non-conservative Maxwell model type. Such
consideration may give place to results of a central limit theorem
type for non classical statistical equilibrium stationary states.
 In the classical case of Maxwell Molecules models  for energy
 conservation (that is for $\mu(1)=0$) with bounded initial energy,
these connections has been fully established recently by
\cite{CaCaGa1, CaCaGa2}, where in this case the asymptotic states
are Maxwellian equilibrium distributions.
\end{remark}

 %%%%%%%%%%%%%%%%%%%%%%%%%%%%%%%%%%%%%%%%%%%%%%%%%%%
%%%%%%%%%%%%%%%  Section 8 %%%%%%%%%%%%%%%%%%%%%%%%%
%%%%%%%%%%%%%%%%%%%%%%%%%%%%%%%%%%%%%%%%%%%%%%%%%%%
\vskip.5in

 \section{Properties of self-similar solutions}

The goal of this Section  is to apply the general theory (in particular,
Theorem~\ref{thm7.2}) to the particular case of the multi-linear
 operators $\Gamma$ considered in Section 4, where their
corresponding  spectral function $\mu (p)$
satisfies \eqref{eq7.11}, \eqref{eq5.3} and its behavior corresponds to
 Fig.1.  We also show that $p_0>1$ is a necessary
 condition for self-similar asymptotics, for $p_0$ being the unique minimum of
 the spectral function,
 that is  $\mu(p_0)= \min_{p>0}\mu(p)$.

In addition, Theorem~\ref{thm8.3} establishes sufficient  conditions
for which self-similar solutions of problem \eqref{eq7.1} will lead
to well defined
 self-similar solutions (distribution functions) of the
original problem after taking the inverse Fourier transform.

\medskip

 We consider the integral equation \eqref{eq7.2} written as
 \begin{equation}\label{eq8.1}
 w= \Gamma_\mu (w) = \int_0^1 d\tau\, g(x\tau^\mu)\ ,\qquad
 g = \Gamma (w)\ ,\ \mu \in\real\ .
 \end{equation}
 First we establish two properties of $w(x)$ that are independent of
the  specific form \eqref{eq4.3} of $\Gamma$.

 \begin{lem}\label{lem8.1}
 $\quad$
 \begin{itemize}
 \item[{{\bf i-}}]  If there exist a closed subset $U' \subset U$ of the unit ball
 $U$  in $B$, as given in \eqref{eq4.9},  such that
$\Gamma _{\mu_*} (U')\subset U'$
 for any $\mu_* \in\real$, and for some function $w_0 \in U'$ the conditions
 of Theorem~\ref{thm7.1} are satisfied,  then $w\in U'$, where $w$ is constructed
 by the iterative scheme
 as defined in \eqref{eq7.3}.

\item[{{\bf ii-}}] If the conditions of Theorem~\ref{thm7.1} for $\Gamma$ are
 satisfied and, in addition, $\Gamma (1) =1$,
 \end{itemize}
 then the solution $w_* =1$
 of Eq.~\eqref{eq8.1} is unique in the class of functions $w(x)$
 satisfying the condition
 \begin{equation}\label{eq8.2}
 w(x) =1 + O (x^p)\ ,\qquad \mu (p) <\mu_*\ .
 \end{equation}
 \end{lem}

 \begin{proof}
 The first statement follows from the iteration scheme \eqref{eq7.3} with
 $w_0 \in U'$.
Then, by assumption {\bf i}, $w_0 \in U'$ for all integer $n \geq 1$, and
$w_n \to w\in U'$.
The second statement follows from the obvious fact that $w_* =1$
satisfies Eq.~\eqref{eq8.1} with any $\mu_* \in\real$ provided
$\Gamma (1) =1$ and from the uniqueness of $w(x)$ stated in
Theorem~\ref{thm7.1}.
This completes the proof.
\end{proof}

The statement {\bf ii} can be interpreted as a necessary condition for
existence of non-trivial ($w\ne \text{const.}$) solutions of Eq.~\eqref{eq8.1}:

\noindent {\it if there exists a non-trivial solution $w(x)$ of Eq.~\eqref{eq8.1} for any
$\mu_*$, where
$\Gamma (1) =1$, such that
\begin{equation}\label{eq8.3}
\|w\| =1\ ,\qquad w= 1 + O (x^p)\ ,\quad p>0\, ,\ \
\ \ \ \text{{\it then}}\ \ \ \ \ \ \  \ \ \ \mu_* \le \mu(p)\, .
\end{equation}  }

We recall that $\mu (p)$ satisfies the inequality $\mu (p) \ge \mu(p_0)=\min_{p>0}\mu(p)$
(see Fig.1).

Moreover, if $p \ge p_0$ (provided $p_0 <\infty$) in Eqs.~\eqref{eq8.3}, then all
solutions of Eq.~\eqref{eq8.1} are trivial, for any  $\mu_* >\mu (p_0)$,
 since $\mu(p)$ is increasing in $(p_0,\infty)$ .

On the other hand, possible solutions with $\mu \le \mu(p_0)$ (even
if they exist) are irrelevant for the problem~\eqref{eq4.6} since they
have an  empty domain of attraction (Lemma~\ref{lem6.1}).

Therefore we always assume below that $\mu >\mu (p_0)$ and,
consequently, $p \in (0,p_0)$ in Eq.~\eqref{eq8.3}.

\medskip

Let us consider now the specific class \eqref{eq4.3}--\eqref{eq4.3bis}
 of operators $\Gamma$,
with functions $u (x)$ satisfying the condition $u(0)=1$.
That is, $u(0,t) =1$ for the solution $u(x,t)$ of the
problem \eqref{eq4.6}.

Since the operators \eqref{eq4.3} are invariant under dilation
 transformations  \eqref{eq6.2} (property {\bf (c)}, Section 6),
 the problem \eqref{eq4.6} with the initial condition $u_0(x)$
satisfying
\begin{equation}\label{eq8.4}
u(0) =1\ ,\quad \|u_0\| = 1\ ;\qquad u_0 (x) =1 - \beta x^p + \cdots,\quad
x\to 0\ ,
\end{equation}
can be always reduced to the case $\beta =1$ by the transformation
$x' = x\beta^{1/p}$.

Moreover, the whole class of operators \eqref{eq4.3}, with different
kernels $A_n (a_1,\ldots,a_n)$, $n = 1,2,\ldots$, is invariant under
transformations $\tilde x = x^p$, $p>0$.
The result of such transformation acting on  $\Gamma$ is another operator
$\widetilde\Gamma$ of the same class \eqref{eq4.3} with kernels
$\tilde A_n (a_1,\ldots,a_n)$.

Therefore, we fix the initial condition \eqref{eq8.4} with $\beta
=1$ and transform the function \eqref{eq8.4} and the Eq.
\eqref{eq4.6} to new variables $\tilde x = x^p$. Then, we omit the
tildes and reduce the problem \eqref{eq4.6}, with initial condition
\eqref{eq8.4} to the case $\beta =1$, $p=1$. We study this case,
which correspond to finite energy, in detail and formulate afterward
the results in terms of initial variables.

\medskip

Next, our goal now is to apply the general theory (in particular,
Theorem~\ref{thm7.2} and  criterion \eqref{eq7.13a}) to the
particular case where the initial data $u_0(x)$ satisfies,
%we assume a bit more about the asymptotics of the initial data
 for small  $x$
\begin{equation}\label{eq8.5}
\|u_0\| =1\ ,\quad u_0 (x) = 1-x + O(x^{1+\ep})\ ,\quad x\to 0\ ,
\end{equation}
with some $\ep>0$.

We also assume that the spectral function $\mu (p)$ given by
\eqref{eq7.11}, \eqref{eq5.3}, which  corresponds to one of the four
cases shown on Fig.1 with  a unique minimum achieved at $p_0 >1$.

We shall prove that any solution constructed by the iteration scheme
of  theorem \eqref{thm7.1} with $\mustar=\mu(1)$, satisfying also the
asymptotics from
theorem \eqref{thm7.2} is controlled from
below by $e^{-x}$
 and that $\lim_{x\to\infty} w(x) =0$, as well as other properties
 with a significant meaning.

To this end, let us take a typical function $u_0 = e^{-x}$
satisfying \eqref{eq8.5}  and apply the criterion \eqref{eq7.13a},
 from Theorem~\ref{thm7.2} or, equivalently, look for such   $p>0$
 for which
 \eqref{eq7.13a} is satisfied. That is, find possible values of $p>0$
such that
\begin{equation}\label{eq8.6}
\Gamma_{\mu (p)} (e^{-x}) - e^{-x} = 0 (x^{p+\ep})\, ,
\end{equation}
in the notation of Eq.~\eqref{eq8.1}.

It is important to observe that now the spectral
function $\mu (p)$ is closely
connected with the operator $\Gamma$ (see Eqs.~\eqref{eq7.11}
and \eqref{eq5.3}), since this was not assumed in the
general theory of Sections~4--7.
This connection leads to much more specific results, than, for
example, the general Theorems~\ref{thm7.1}, \ref{thm7.2}.

 Then,  in order to study the properties of self-similar solutions and its asymptotics
to problem \eqref{eq7.1} for $p_0>1$, and consequently, for
\begin{equation}\label{eq8.9}
\mu (p) \ge \mu (p_0) > - \frac1{p_0} > -1\, ,
\end{equation}
 we first investigate the structure of
 $\Gamma_\mu (e^{-x})$ for any $\mu>-1$.
Its  explicit formula  reads
\begin{equation}\label{eq8.7}
\Gamma_\mu (e^{-x} )
= \sum_{n=1}^N \alpha_n \int_{\real_+^n} da_1\ldots da_n A_n
(a_1,\ldots,a_n) I_\mu \Big[ x \sum_{k=1}^n a_k\Big]\ ,
\end{equation}
where
\begin{equation}\label{eq8.8}
I_\mu (y) = \int_0^1 d\tau\, e^{-y\tau^\mu}\ ,\ \ \quad
\mu\in\real\, , \  y>0\, ,\qquad \qquad \sum_{n=1}^N \alpha_n =1\, .
\end{equation}

Hence, in order to find  some properties of $\Gamma_\mu (y)$,
for any  $\mu >-1$,
we  prove the following lemma.

\begin{lem}\label{lem8.2}
If $\mu >-1$,  $y\ge 0$, then $0<I_\mu (y) \le1$, and
\begin{equation}\label{eq8.10}
I_\mu (y) = e^{-y} \left( 1 + \frac{\mu y}{1+\mu}\right)
+ \frac{\mu^2}{1+\mu} \, r_\mu (y)\ ,
\end{equation}
where $0\le r_\mu (y) \le B(y,\mu)$ with
\begin{equation}\label{eq8.11}
B(y,\mu)  = \begin{cases}
 {y^2}{(2\mu +1)}^{-1} \ \qquad &\text{ if }\ \mu > -\frac12\, ,\\
  2y^2 \left( -\ln y +y\right)
 \qquad &\text{ if }\ \mu = -\frac12\, ;\\
 \frac{\Gamma(2-|\mu|^{-1})} {|\mu|}  \ y^{\frac1{|\mu|}}
\ &\text{ if }\ \mu \in \left(-1,-\frac12\right)\ .
\end{cases}
\end{equation}
\end{lem}

\begin{proof}
We consider the integral $I_\mu (y)$, integrate twice by parts and obtain
Eq.~\eqref{eq8.10} with
$$r_\mu (y) = y^2 \int_0^1 \, e^{-y\tau^\mu} \tau^{2\mu}\, d\tau\ .$$
If $\mu > -1/2$ then the estimate \eqref{eq8.11} is obvious.
Otherwise we transform $r_y(y)$ into the form
\begin{equation*}
r_\mu (y) = \frac1{|\mu|} y^{\frac1{|\mu|}} \int_y^\infty  e^{-\tau}
\tau^{1-\frac1{|\mu|}}\, d\tau\ ,\qquad \mu \le -\frac12\ ,
\end{equation*}
so that the estimate \eqref{eq8.11} with $\mu \in(-1,-\frac12)$ is also clear.

Finally, in the case $\mu = -\frac12$ we obtain
\begin{equation*}
r_{-1/2} (y) = 2y^2 E(y)\ ,\qquad \text{with}\ \ E(y) =
\int_y^\infty \frac{ e^{-\tau}}{\tau} \, d\tau\ .
\end{equation*}
Hence, rewriting
\begin{equation*}
\begin{split}
E(y) & = \int_1^\infty \frac{ e^{-\tau}}{\tau}\, d\tau + \int_y^1\,
\frac{d\tau}{\tau}
+ \int_y^1 (e^{-\tau} -1)\, \frac{d\tau}{\tau}  \\
\noalign{\vskip6pt} & = -\ln y  + \int_0^y (1-e^{-\tau})\,
\frac{d\tau}{\tau} + C\ ,
\end{split}
\end{equation*}
where $C$ is the well-known integral
\begin{equation*}
C = \int_1^\infty \frac{e^{-\tau}}{\tau}\, d\tau + \int_0^1
 (e^{-\tau}-1) \, \frac{d\tau}{\tau} = \int_0^\infty  e^{-\tau} \ln
\tau\, d\tau = -\gamma\ ,
\end{equation*}
with  $\gamma = \Gamma' (1) \simeq 0.577$ is the Euler constant.
Therefore, since  $C<0$, and the function $(1-e^{-\tau})\tau^{-1}$
takes the value $1$ at the origin and is positive decreasing in
$(0,y)$ for any $y>0$, then the term $E(y) $ is estimated by
\begin{equation*} E(y) \le - \ln y +
y\ .
\end{equation*}
Hence, we obtain the estimate \eqref{eq8.11} for $\mu = -1/2$ and
the proof is completed.
 \end{proof}

Hence, we  can characterize now the possible  values of $p>0$ for which
 criterion \eqref{eq7.13a} holds, so Theorem~\ref{thm7.2} yields the
self-similar asymptotics. We state and prove this characterization
in the following corollary.

 \begin{corollary}\label{cor6}
 Whenever
 $p_0> 1$ for  $\mu(p_0)=\min_{_{p>0}}\mu(p)$,
 the condition \eqref{eq8.6} is fulfilled if and only if
 $p\le 1$ and, therefore,
 $\mu (p) \ge \mu (1)$ .
\end{corollary}

\begin{proof}
From the previous Lemma we obtain
 \begin{equation*}
 I_\mu (y) = 1-\frac{y}{1+\mu} + O (y^{1+\ep})\ ,\qquad
 \ep >0\ ,\ y\to 0\ ,
 \end{equation*}
 provided $\mu > -1$.
 Therefore
 \begin{equation*}
 \Gamma_\mu (e^{-x}) - e^{-x} = \theta (\mu) x + O (x^{1+\ep})\ ,
 \end{equation*}
 where
 \begin{equation*}
 \theta (\mu ) = 1 - \frac1{1+\mu} \sum_{n=1}^N \alpha_n
 \int_{\real_+^n} da_1\ldots da_n A_n (a_1,\ldots, a_n)
 \sum_{k=1}^n a_k\ .
 \end{equation*}
 We recall that kernels $A_n (a_1,\ldots,a_n)$, $n=1,\ldots,N$,
 are assumed to be symmetric functions of their arguments.
 Then
 \begin{equation}\label{eq8.12}
 \theta (\mu ) = 1 - \frac1{1+\mu} \lambda (1)\ ,\qquad
 \lambda (p) = \int_0^\infty da\, K(a) a^p\ ,
 \end{equation}
 where $K(a)$ is given in Eqs.~\eqref{eq5.3}.
 Recalling  the definition of $\mu (p)$ in  \eqref{spec-func},
 we obtain
 \begin{equation}\label{eq8.13}
 \Gamma_{\mu (p)} (e^{-x}) - e^{-x} =
 \theta [\mu (p)] x + O (x^{1+\ep})\ ,
 \end{equation}
 where
 \begin{equation}\label{eq8.14}
 \theta [\mu (p)] = \frac{\mu (p) - \mu (1)}{1+\mu (p)} \ ,\qquad
 0 < p < p_0\ .
 \end{equation}

It follows from Eqs.~\eqref{eq8.13}, \eqref{eq8.14} that  the
condition \eqref{eq8.6} is fulfilled
 if and only if  $p\le 1$.
On the other hand, it was assumed above that $p_0>1$, therefore $\mu
(p) \ge \mu (1)$ for such values of $p$. Thus, Corollary~\ref{cor6}
is proved.

 \end{proof}

\medskip

 In addition, by Lemma~\ref{lem8.1}~{[ii]},   if $\mustar > \mu (1)$
 then the iteration scheme
 \begin{equation}\label{eq8.15}
 w_{n+1} = \Gamma_{\mustar} (w_n)\ ,\qquad w_0 = e^{-x}\ ,\ n= 0,1,\ldots,
 \end{equation}
  converges to the trivial solution $w=1$.
 Hence, the only nontrivial case corresponds to $\mustar  = \mu (1)$
 in Eqs.~\eqref{eq8.15}.

 Hence, according to Theorem~\ref{thm7.1}, $w_n (x) \to w(x)$, where
 $w(x)$ is continuously differentiable function on $[0,\infty)$.

 On the other hand,
 \begin{equation}\label{eq8.16}
 w(x) = w_0 (x) + O(x^{1+\ep})
 \end{equation}
 and therefore $w'(0) = w'_0(0) = -1$.
 Since this condition is not fulfilled for $w(x) =1$,
  we indeed obtain a non-trivial solution of Eq.~\eqref{eq8.1}
 with $\mustar = \mu (1)$.

 Even though we would obtain the same result starting the iterations
 \eqref{eq8.15} from any initial function satisfying Eqs.~\eqref{eq8.5},
 the specific function $w_0 = e^{-x}$ has, however, some advantages
 since it gives some additional information about the properties of $w(x)$.

 {From} now on we assume that $\mustar = \mu (1)$ in Eqs.~\eqref{eq8.15}
 and denote
 \begin{equation}\label{eq8.17}
 w(x) = \lim_{n\to\infty} w_n(x)\ ,\qquad x\ge 0\ .
 \end{equation}
 Then, by Theorem~\ref{thm7.1}, $w\in C_1 (\real_+)$ and satisfies
 the equation
 \begin{equation}\label{eq8.18}
 \mustar \, x\, w'(x) + w(x)
 = \Gamma (x)\ ,\qquad \mustar = \mu (1)\ .
 \end{equation}
 The differentiability of $w(x)$ was proved in Theorem~\ref{thm7.1}
 only for $\mu \ne0$, but the proof can be easily extended to the
 case $\mu =0$ since $w_0 = e^{-x}$ in Eqs.~\eqref{eq8.15} has a  bounded
 and continuous derivative.

 From \eqref{eq8.15} and \eqref{eq8.16}  it is clear that the limit function $w$
 satisfies
 \begin{equation}\label{eq8.19}
 0\le w(x) \le 1\ ,\qquad w(0) =1\ ,\ w'(0) = -1\ ;
 \end{equation}
and, by considering a sequence of derivatives in Eqs.~\eqref{eq8.15},
 it is easy to see that
 \begin{equation}\label{eq8.20}
 w'(x) \le 0\ ,\qquad |w'(x)| \le 1\ .
 \end{equation}
Then,  estimates from Theorem~\ref{thm7.1} and Lemma~\ref{lem8.2}
 yield that
 \begin{equation}\label{eq8.21}
 w(x) = e^{-x} + O (x^{\pi (\mustar)})\ ,
 \end{equation}
 where
 \begin{equation}\label{eq8.22}
 \pi(\mustar) = \begin{cases}
2 \qquad \ \ &\text{ for  } \ \mustar >-\frac12\ ,\\
  2-\ep \text{ with any }\ep> 0\ &\text{ for  }\  \mustar =-\frac12\ ,\\
 \frac1{|\mustar|} \ \ \ &\text{ for }\  -1 <\mustar < -\frac12\ .
 \end{cases}
 \end{equation}

 Hence, we collect all essential properties of $w(x)$ in the following
 statement.

 \begin{thm}\label{thm8.3}
 The limiting function $w(x)$ constructed in
 \eqref{eq8.15} satisfies Eq.~\eqref{eq8.1}
 with $\mu = \mu (1)$ and Eqs.~\eqref{eq8.18}, where $\Gamma$ is
 given in Eqs.~\eqref{eq4.3}, $\mu (p)$ is defined in Eqs.~\eqref{eq7.11},
\eqref{eq5.3}.
 The conditions \eqref{eq8.19}, \eqref{eq8.20}, \eqref{eq8.21} are
 fulfilled for $w(x)$.
 Moreover
 \begin{equation}\label{eq8.23}
 1 \ge w (x) \ge e^{-x}\ ,\qquad
 \lim_{x\to\infty} w(x) = 0\ ,
 \end{equation}
 and there exists a generalized non-negative function
 $R(\tau)$, $\tau \ge 0$, such that
 \begin{equation}\label{eq8.24}
 w(x) = \int_0^\infty d\tau\, R(\tau) e^{-\tau x}\ ,\qquad
 \int_0^\infty d\tau\, R(\tau)
 = \int_0^\infty d\tau\, R(\tau) \tau =1\ .
 \end{equation}
 \end{thm}

 \begin{proof}
 It remains to prove \eqref{eq8.23} and  \eqref{eq8.24}.
In fact condition \eqref{eq8.23} means that  $e^{-x}$ is a {\sl
barrier function} to the solutions of problem \eqref{eq8.1}  with
$\mu=\mu(1)$.

 First we note that Eq.~\eqref{eq8.1} is obtained as the integral
 form of Eq.~\eqref{eq7.1}.
Then, the identity
\begin{equation*}
\mu  \, x\, v' (x) + v (x) = \Gamma (v) + \Delta (x)\, ,
\end{equation*}
where
\begin{equation}\label{eq8.25}
\Delta (x) = \mu  \, x\, v' (x) + v(x) - \Gamma (v)\, ,
\end{equation}
is fulfilled for any function $v(x)$, and the integral form of this
identity reads
\begin{equation*}
v(x) = \Gamma_\mu (v) + \int_0^1   \Delta (x\tau^\mu)\, d\tau \ .
\end{equation*}
Hence, if $\Delta (x) \le 0$ then $v\le \Gamma_\mu (v)$ and
vice-versa.

 We intend to prove that $\Delta (x) \le 0$ for $v=
e^{-x}$. If so, then $w_{n+1}(x) \ge w_n (x)$ at any $x\ge 0$ in the
sequence \eqref{eq8.15} generated by the corresponding iteration
scheme with $w_0=e^{-x}$, and obviously $w(x) \ge e^{-x}$.

Indeed, by substituting $v = e^{-x}$ in Eqs.~\eqref{eq8.25} we
obtain, for $\mu = \mu (1)$,
\begin{equation*}
\Delta (x) = \sum_{n=1}^\infty \alpha_n \Delta_n (x)\ ,
\end{equation*}
where using \eqref{eq8.8},
\begin{gather*}
\Delta_n (x) = \int_{\real_+^n} da_1,\ldots, da_n A_n
(a_1,\ldots,a_n) P \bigg( x,\sum_{k=1}^n a_k\bigg)\ ,\text{\ \ \ with} \\
\noalign{\vskip6pt}
P(x,s) = e^{-x} [1-(s-1)x] - e^{-sx}\le 0\ .
\end{gather*}
We note that $P(x,s) \le 0$ for any real $s$ and $x$, since $e^y \le
1 +y$ for any real $y$.  Then $\Delta_n (x)\le 0 $, and so also
$\Delta (x)\le 0 $. Hence, the inequality in \eqref{eq8.23} is
proved.

In order to prove the limiting identity \eqref{eq8.23} we denote
\begin{equation*}
w_\infty = \lim_{x\to \infty} w(x)\ .
\end{equation*}
Such limit exists since $w(x)$ is a monotone function. {From}
theorem~\ref{thm8.3}, the nice properties of $w(x)$ allow to take
the limit in both sides of Eq.~\eqref{eq8.1}. Then
\begin{equation*}
w_\infty = \sum_{n=1}^N \alpha_n w_\infty^n\ ,\qquad
\sum_{n=1}^N \alpha_n = 1\ ,\quad
\alpha_n \ge 0\ ,
\end{equation*}
and therefore we obtain
\begin{equation*}
\sum_{n=2}^\infty \alpha_n w_\infty (1-w_\infty^{n-1}) =0\ .
\end{equation*}
This equation has  just two non-negative roots: $w_\infty =0$ and
$w_\infty =1$. The root $w_\infty =1$ is possible only if $w(x)=1$
for all real $x$. Since by \eqref{eq8.16}  this is not the case,
then $w_\infty =0$.

It remains to prove the integral representation \eqref{eq8.24}. In
order to do this we denote by $U'$ the set of Laplace transforms of
probability measures in $\real_+$, i.e., $u\in U'$ if there exists a
generalized function $F(\tau) \ge0$ such that
\begin{equation*}
u(x) = \int_0^\infty d\tau\, F(\tau) e^{-x{\tau}}\  ,\qquad
\int_0^\infty d\tau\, F(\tau) =1\ .
\end{equation*}
Then $e^{-x}\in U'$ (with $F = \delta (\tau-1)$) and it is easy to
check that $\Gamma_\mu (U') \subset U'$ for any real $\mu$. On the
other hand, the set  $U'$ is closed with respect to uniform
convergence in $\real_+$ (see, for example, \cite{Fe}). Thus,
according to Lemma~\ref{lem8.1}~[i], $w\in U'$. On the other hand,
 it is already known from \eqref{eq8.19} that $w'(0) =-1$. Hence, the
corresponding function $R(\tau)$ has a unit first moment \cite{Fe}.
This completes the proof of Theorem~\ref{thm8.3}.
\end{proof}

The integral representation \eqref{eq8.24} is important for the
properties of the corresponding distribution functions satisfying
Boltzmann-type equations. Now it is easy to return to initial
variables with $u_0$ given in Eq.~\eqref{eq8.4} and to describe the
complete picture of the self-similar relaxation for the problem
\eqref{eq4.6}.

\vskip.5in
%%%%%%%%%%%%%%%%%%%%%%%%%%%%%%%%%%%%%%%%%%%%%%%%%%%%%%%%%
 %%%%%%%%%%%%%%%%%%% Section 9  %%%%%%%%%%%%%%%%%%%%%%%%%%%%
%%%%%%%%%%%%%%%%%%%%%%%%%%%%%%%%%%%%%%%%%%%%%%%%%%%%%%%%%%%%%%

 \section{Main results for Fourier transformed Maxwell models\\
 with multiple interactions}

 We consider the Cauchy problem \eqref{eq4.6} with a fixed operator
  $\Gamma$ \eqref{eq4.3} and study the time evolution of $u_0(x)$
  satisfying the conditions
  \begin{equation}\label{eq9.1}
  \|u_0\| = 1\ ; \quad u_0 = 1 - x^p + O(x^{p+\ep})\ ,\   x\to 0\ ,
  \end{equation}
  with some positive $p$ and $\ep$.
  Then there exists a unique classical solution $u(x,t)$ of the
  problem \eqref{eq4.6}, \eqref{eq9.1} such that, for all $t\ge0$,
  \begin{equation}\label{eq9.2}
  \|u(\cdot,t) \| =1\ ;\qquad
  u(x,t) =1 + O(x^p)\ ,\ x\to 0\ .
  \end{equation}
  We explain below the simplest way to analyze this solution,
  in particular in the case of self-similar asymptotics.

 \step{Step 1}
  Consider the linearized operator $L$ given in
  Eqs.~\eqref{eq5.2}--\eqref{eq5.3} and construct the spectral
  function $\mu (p)$ given in Eq.~\eqref{eq7.11}.
  The resulting $\mu (p)$ will be of one of four kinds
  described qualitatively on Fig.1.

  \step{Step 2}
  Find the value $p_0 >0$ where  the minimum (infimum)
  of $\mu (p)$ is achieved.
  Note that $p_0 =\infty$ just for the case described on Fig.1~(a),
  otherwise $0<p_0<\infty$.
  Compare $p_0$ with the value $p$ from Eqs.~\eqref{eq9.1}.

  If $p<p_0$ then the problem \eqref{eq4.6}, \eqref{eq9.1} has a
  self-similar asymptotics (see below).

  The above consideration shows that two different cases are
  possible:
\begin{itemize}
  \item[(1)]~$p\ge p_0$ provided $p_0<\infty$;
\item[(2)]~$0<p< p_0$, that is the spectral function $\mu(p)$ in monotone
decreasing for all $0<p<p_0$.
\end{itemize}

  In case (1) a behavior of $u(x,t)$ for large $t$ may depend
  strictly on initial conditions.
  The only general conclusion that can be drawn for the initial data
  \eqref{eq9.1} with $p \ge p_0$ is the following:
  \begin{equation}\label{eq9.3}
  \lim_{t\to \infty} u(xe^{-\mu t}, t) = 1\ ,\qquad x\ge 0\ ,
  \end{equation}
  for any $\mu >\mu (p_0)$.
  This follows from Lemma~\ref{lem6.4} with $u^{(1)} =u$,
  $u^{(2)} =1$ and from the fact that any such function $u_0(x)$
  satisfies the condition
  \begin{equation*}
  u_0 =1 + O(x^{p_0})
  \end{equation*}

  Case (2) with $0<p<p_0$ in Eqs.~\eqref{eq9.1} is more
  interesting.
  In this case (assume that $p\in (0,p_0)$ is fixed) there exists a
  unique self-similar solution
  \begin{equation}\label{eq9.4}
  u_s (x,t) = \psi (xe^{\mu (p)t} )
  \end{equation}
  satisfying Eqs.~\eqref{eq8.1} at $t=0$.
  We again use Lemma~\ref{lem6.4} with $u_{1} =u$ and
  $u_2 = u_s$ and obtain for the solution $u(x,t)$ of the problem
  \eqref{eq4.6}, \eqref{eq9.1}:
  \begin{equation}\label{eq9.5}
  \lim_{t\to\infty} u (xe^{-\mu t},t) =
  \begin{cases}
  1&\text{if }\ \mu >\mu (p)\\
  \psi (x)&\text{if }\ \mu = \mu (p)\\
  0&\text{if }\ \mu (p) > \mu > \mu (p +\delta)\ ,
  \end{cases}
  \end{equation}
  with sufficiently small $\delta >0$.
  The third equality follows from the fact that
  \begin{equation*}
  u_0(x) - \psi (x) = O(x^{p+\ep})
  \end{equation*}
  and from the equality (see Eqs.~\eqref{eq8.23})
  \begin{equation*}
  \lim_{x\to\infty} \psi (x) =0\ .
  \end{equation*}

  We note that $\psi (x) = w(x^p)$, where $w(x)$ has all
  properties described in Theorem~\ref{thm8.3}.
  The equalities \eqref{eq9.5} explain the exact meaning of
  the approximate identity,
  \begin{equation}\label{eq9.6}
  u(x,t) \approx \psi ( xe^{\mu (p)t})\ ,\qquad
  t\to \infty\ ,\ xe^{\mu (p)t} = \text{const.}\ ,
  \end{equation}
  that we call self-similar asymptotics.
  We collect the results in the following statement.

  \begin{prop}\label{prop9.1}
  The solution $u(x,t)$ of the problem \eqref{eq4.6}, \eqref{eq9.1},
  with $\Gamma$ given in Eqs.~\eqref{eq4.3}, satisfies either one
of  the following
  limiting identities:
 \begin{itemize}
 \item[{\bf (1)}] ~Eq.~\eqref{eq9.3} if $p\ge p_0$ for the initial
data \eqref{eq9.1}\, ,
   \item[{\bf (2)}] ~Eqs.~\eqref{eq9.5} provided $0<p<p_0$.
 \end{itemize}
The  convergence in Eqs.~\eqref{eq9.3}, \eqref{eq9.5} is
uniform on
  any bounded interval $0\le x\le R$, and
 \begin{equation*}
u ( xe^{\mu (p)t},t) -\psi(x) = O(x^{p+\ep}) e^{-\beta(p,\ep)t}\, ,\qquad
\beta(p,\ep)= (p+\ep)\,(\mu(p) -\mu(p+\ep) \,),
\end{equation*}
for   $0<p<p_0$ and $0<\ep<p_0-p$. \end{prop}

  \begin{proof}
  It remains to prove the last statement.
  It follows in both cases from the estimate \eqref{eq6.21} for
the  remainder term in Lemma~\eqref{lem6.4}.
   This completes the proof.
  \end{proof}

  It is interesting that our considerations are the same for both positive
  and negative values of $\mu (p)$.
  There are, however, certain differences if we want to consider the
  ``pure'' large time asymptotics, i.e., the limits \eqref{eq9.3},
  \eqref{eq9.5} with $\mu=0$.
  Then we can conclude that
  \begin{equation*}
  \begin{split}
 \text{\bf (1)}\quad & \lim_{t\to\infty} u(x,t)  = 1\ \ \ \ \ \ \ \
\text{if}\  \  p\ge p_0\text{ and } \mu (p_0)<0\, , \
  \text{ or } 0 < p< p_0\text{ and } \mu (p) <0\ ;\\
 \text{\bf (2)}\quad & \lim_{t\to\infty} u(x,t)  = \psi (x) \ \
\text{ if }\ \  0<p<p_0\text{ and }
  \mu (p)=0\ .
  \end{split}
  \end{equation*}

  It seems probable that $u(x,t)\to0$ for large $t$ in all other cases, but
  our results, obtained on the basis of Lemma~\ref{lem6.4}, are not
  sufficient to prove this.

\begin{remark} We mention that the self-similar asymptotics becomes more
  transparent in logarithmic variables
  \begin{equation*}
  y = \ln x\ ,\quad u(x,t) = \hat u(y,t)\ ,\quad \psi (x,t) = \hat\psi(y,t)
  \end{equation*}
  Then Eq.~\eqref{eq9.6} reads
  \begin{equation}\label{eq9.7}
  \hat u(y,t) \approx \hat\psi (y + \mu (p)t)\ ,\qquad
  t\to \infty\ ,\quad y+\mu (p)t = \text{const. },
  \end{equation}
  i.e., the self-similar solutions are simply nonlinear waves
  (note that $\psi (-\infty) =1$, $\psi (+\infty) =0$)
  propagating with constant velocities $c_p = -\mu (p)$ to
  the right if $c_p >0$ or to the left if $c_p <0$.
  If $c_p >0$ then the value $u(-\infty,t)=1$ is transported
  to any given point $y\in\real$ when $t\to\infty$.
  If $c_p <0$ then the profile of the wave looks more
  natural for the  functions $\tilde u =1-\hat u$, $\tilde\psi=1-\psi$.

Thus, Eq.~\eqref{eq4.6} can be considered in some sense as
  the equation for nonlinear waves.
  The self-similar asymptotics \eqref{eq9.7}
  means a formation of the traveling wave with a universal
  profile for a broad class of initial conditions.
 This is a purely non-linear phenomenon, it is easy to see
  that such asymptotics cannot occur in the particular case
  ($N=1$ in Eqs.~\eqref{eq4.3}) of the linear operator $\Gamma$.

\end{remark}

\vskip.5in
%%%%%%%%%%%%%%%%%%%%%%%%%%%%%%%%%%%%%%%%%%%
 %%%%%%%%%%%%%%%%%%% section 10 %%%%%%%%%%%%%%%%%%%%
%%%%%%%%%%%%%%%%%%%%%%%%%%%%%%%%%%%%%%%%%%%%%%%%%
 \section{Distribution functions, moments and power-like tails}

 We have described above the general picture of the behavior of the solutions
  $u(x,t)$ to the problem \eqref{eq4.6}, \eqref{eq9.1}.
  On the other hand, Eq.~\eqref{eq4.6} (in particular, its special
  case \eqref{eq3.2}) was obtained as the Fourier transform of
  the kinetic equation.
  Therefore we need to study in more detail the corresponding
  distribution functions.

 We assume in this section that $u_0(x)$ in the problem
 \eqref{eq4.6} is an isotropic characteristic function of a
 probability measure in $\real^d$, i.e.,
 \begin{equation}\label{eq10.1}
 u_0(x) = \F [f_0] = \int_{\real^d} dv\, f_0 (|v|) e^{-ik\cdot v}\ ,
 \qquad k\in \real^d\ ,\ x= |k|^2\ ,
 \end{equation}
 where $f_0$ is a generalized positive function normalized
 such that $u_0(0)=1$ (distribution function).
 Let $U$ be a closed unit ball  in the $B=C(\real_+)$ as defined
 in \eqref{eq4.10}.
 Then,  as was already mentioned at the end of Section~5, the
 set $U'\subset U$ of isotropic characteristic functions is a
 closed convex subset of $U$.
 Moreover, $\Gamma (U') \subset U'$ if $\Gamma$ is defined
 in Eqs.~\eqref{eq4.3}.
 Hence, we can apply Lemma~\ref{lem5.3} and conclude that there
 exists a distribution function $f(v,t)$, $v\in\real^d$, satisfying
 Eq.~\eqref{eq2.1}, such that
 \begin{equation}\label{eq10.2}
 u(x,t) = \F [f(\cdot,t)]\ ,\qquad x = |k|^2\ ,
 \end{equation}
 for any $t\ge 0$.

A  similar conclusion can be obtain if  we assume the Laplace
 (instead of Fourier) transform in Eqs.~\eqref{eq10.1}.
 Then there exists a distribution function $f(v,t)$, $v>0$,
 such that
 \begin{equation}\label{eq10.3}
 u(x,t) = \L [f(\cdot,t)] = \int_0^\infty dv\, f(v,t) e^{-xv}\ ,\qquad
 u(0,t) = 1\ ,\ \ x\geq 0\,,\ \ t\ge 0\ ,
 \end{equation}
 where $u(x,t)$ is the solution of the problem \eqref{eq4.6}
 constructed in Theorem~\ref{thm5.2} and Lemma~\ref{lem5.3}.

We remind  the reader that the point-wise convergence $u_n (x) \to
u(x)$, $x\ge0$, where $\{u_n,\, n=1,2,\ldots\}$ and $u$ are
characteristic functions (or Laplace transforms) is
 sufficient for the weak convergence of the corresponding probability
 measures \cite{Fe}.  Hence, all results of pointwise convergence
 related to self-similar asymptotics
can be easily re-formulated in corresponding  terms for the related
distribution functions (or, equivalently, probability measures).

 The approximate equation \eqref{eq9.6} in terms of
 distribution functions \eqref{eq10.2} reads
 \begin{equation}\label{eq10.4}
 f(|v|,t) \simeq e^{-\frac{d}2 \mu(p)\, t} F_p (|v| e^{-\frac12 \mu(p)\,t})\ ,
 \quad t\to \infty\ ,\ |v| e^{-\frac12 \mu(p)\, t} = \text{const.}\ ,
 \end{equation}
 where $F_p (|v|)$ is a distribution function such that for $x= |k|^2$
 \begin{equation}\label{eq10.5}
 \psi_p (x) = \F [F_p]\ \ ,
 \end{equation}
 with $\psi_p $ given in Eq.~\eqref{eq9.4} (the notation $\psi_p$
 is used in order to stress that $\psi$ defined in \eqref{eq9.4},
  depends on $p$).
 The factor $1/2$ in Eqs.~\eqref{eq10.4} is due to the notation
 $x= |k|^2$.
 Similarly, for the Laplace transform \eqref{eq10.3}, we obtain
 \begin{equation}\label{eq10.6}
 f(v,t) \simeq e^{-\mu (p)t} \Phi_p (ve^{-\mu (p)t})\ ,\qquad
 t\to\infty\ ,\ ve^{-\mu (p)t} = \text{const.}\ ,
 \end{equation}
 where
 \begin{equation}\label{eq10.7}
 \psi_p(x) = \L [\Phi_p]\ .
 \end{equation}

In the space of distributions, the approximate relation $\simeq$ is
weak in the sense of distributions, i.e. the classical approximation
concept of real valued expression obtained after integrating by test
functions.

 The positivity and some other properties of $F_p (|v|)$ follow from
 the fact that $\psi_p (x) = w_p(x^p)$, where $w_p(x)$ satisfies
 Theorem~\ref{thm8.3}.
 Hence
 \begin{equation}\label{eq10.8}
 \psi_p (x) = \int_0^\infty d\tau\, R_p (\tau) e^{-\tau x^p}\ ,
 \qquad \int_0^\infty d\tau\, R_p(\tau)
 = \int_0^\infty d\tau\, R_p (\tau) \tau = 1\ ,
 \end{equation}
 where $R_p (\tau)$, $\tau \ge 0$, is a non-negative generalized
 function (of course, both $\psi_p$ and $R_p$ depend on $p$).

 We stress here that
 if $u_0 (x)$ is a characteristic function given in Eq.~\eqref{eq10.1},
 and  condition \eqref{eq9.1} is fulfilled, then $p\le 1$ in
 Eqs.~\eqref{eq9.1}.
 In addition,
 the case $p>1$ is impossible for the  non-negative initial distribution
$f_0$ in
 Eqs.~\eqref{eq10.1} since
 \begin{equation}\label{eq10.9}
 u' (0) = - C_d \int_{\real^d} dv\, f_0 (|v|) |v|^2\ ,
 \end{equation}
 where $C_d>0$ is a constant factor that depends only on the space
 dimension for the problem.

 Hence, the self-similar asymptotics \eqref{eq10.4} for any
 initial data $f_0\ge 0$ occurs if $p_0 >1$ (see Step~2 at
 the beginning of Section~9).
 Otherwise it occurs for $p\in (0,p_0) \subset (0,1)$.
 Therefore, for any spectral function $\mu (p)$ (Fig.1), the
 approximate relation \eqref{eq10.4} holds for sufficiently
 small $0<p\le 1$.
 It follows from Eq.~\eqref{eq10.9} that
 \begin{equation*}
 m_2 = \int_{\real^d} dv\, f_0 (|v|) |v|^2 <\infty \ \text{ if }\ p=1
 \end{equation*}
 and $m_2=\infty$  if $p<1$.

 Similar conclusions can be made for the Laplace transforms
 \eqref{eq10.3} since in that case
 \begin{equation*}
 u' (0,t) = - \int_0^\infty dv\, f(v,t) v\ ,
 \end{equation*}
 therefore the first moment of $f$ plays the same role as the second
 moment in case of Fourier transforms.

 The positivity of $F(|v|)$ in Eqs.~\eqref{eq10.5} and \eqref{eq10.7}
 follows from the integral representation \eqref{eq10.8} with $p\le1$.
 It is well-known that
 \begin{equation*}
 \F^{-1} (e^{-|k|^{2p}} ) >0\ ,\qquad \L^{-1} (e^{-x^{2p}}) >0
 \end{equation*}
 for any $0<p\le 1$ (the so-called infinitely divisible distributions
 \cite{Fe}).
 Thus, Eqs.~\eqref{eq10.8} explains the connection of the  self-similar
 solutions of generalized Maxwell models with infinitely
 divisible distributions.
 We can use standard formulas for the inverse Fourier (Laplace)
 transforms and denote ($d= 1,2,\ldots$ is fixed)
 \begin{equation}\label{eq10.10}
 \begin{split}
 M_p (|v|)
 & = \frac1{(2\pi)^d} \int_{\real^d} dk e^{-|k|^{2p} +ik\cdot v}\ ,\\
 \noalign{\vskip6pt}
 N_p (v) & = \frac1{2\pi i}\int_{a-i\infty}^{a+i\infty} dx\,
 e^{-x^p +xv}\ ,\qquad 0< p\le 1\ .
 \end{split}
 \end{equation}
 Then the self-similar solutions $F_p$ and $\Phi_p$ (distribution functions)
 given in the right hand sides of Eqs.~\eqref{eq10.6} and  \eqref{eq10.8}
 respectively, satisfy
 \begin{equation}\label{eq10.11}
 \begin{split}
 F_p (|v|) & = \int_0^\infty d\tau\, R_p (\tau) \tau^{-\frac{d}{2p}} M_p
 (|v| \tau^{-\frac1{2p}})\ ,\\
 \noalign{\vskip6pt}
 \Phi_p (v) & = \int_0^\infty d\tau\, R_p (\tau) \tau^{-\frac1p} N_p
 (v\tau^{-\frac1p})\ ,\qquad v \ge 0\ ,\ 0<p\le 1\ .
 \end{split}
 \end{equation}
 That is, they admit an integral representation through infinitely
 divisible distributions.
 Note that $M_1(|v|)$ is the standard Maxwellian in $\real^d$.
 The functions $N_p (v)$ \eqref{eq10.10} are studied in detail
 in the  literature \cite{Fe,Lu}.

 Thus, for given $0<p\le 1$, the kernel $R_p(\tau)$, $\tau\ge0$,
 is the only unknown function that is needed to describe the
  distribution functions $F_p(|v|)$ and $\Phi_p(v)$.
 Therefore we study $R_p(\tau)$ in more detail.

 It was already noted in Section~8 that the general problem
 \eqref{eq8.1}, \eqref{eq8.3}, with given $0<p<p_0$, can be
 reduced to the case $p=1$ by the transformation of variables
 $\tilde x = x^p$.
 We assume therefore that such transformation is already
 made and consider the case $p=1$.
 Then the equation for $R(\tau) = R_1 (\tau)$ can be
 obtained (see Eqs.~\eqref{eq8.24}) by applying the inverse
 Laplace transform to Eq.~\eqref{eq8.18}.
 Then we obtain, with $\mustar=\mu(1)$,
 \begin{equation}\label{eq10.12}
 - \mu(1) \frac{\partial}{\partial \tau} \tau  R(\tau)  + R(\tau)
 = Z(R) = \L^{-1} [\Gamma (w)]\ ,
 \end{equation}
 where (see Eqs.~\eqref{eq4.3})
 \begin{equation*}
 \begin{split}
 &Z(R) = \sum_{n=1}^N\alpha_n Z_n (R)\ ,\qquad
 \sum_{n=1}^N \alpha_n = 1\ ,\ \alpha_n \ge 0\ ,\\
 \noalign{\vskip6pt}
 &Z_n (R) = \int_{\real_+^n} da_1,\ldots, da_n
 \frac{A_n (a_1,\ldots,a_n)}{a_1 a_2 \ldots a_n}
 \prod_{k=1}^n{\!\vphantom{\sum}}^*
 R\Big( \frac{\tau}{a_k}\Big)\ ,\\
\noalign{\vskip6pt}
& \prod_{k=1}^n{\!\vphantom{\sum}}^*
R_k(\tau)  = R_1 * R_2* \ldots * R_n\ ,\qquad
R_1 * R_2 = \int_0^\tau d\tau'\, R_1 (\tau') R_2 (\tau-\tau')\ .
\end{split}
\end{equation*}

Let us denote
\begin{equation}\label{eq10.13}
m_s = \int_0^\infty d\tau\, R(\tau) \tau^s\ ,\qquad s>0\ ,
\end{equation}
then multiply Eq.~\eqref{eq10.12} by $\tau^s$ and obtain
after integration the following equality
\begin{equation}\label{eq10.14}
\begin{split}
&(\mu(1) s+1) m_s = \\
&= \sum_{n=1}^N \alpha_n \int_{\real_+^n} \mkern-18mu
da_1 \ldots da_n A_n (a_1,\ldots,a_n) \int_{\real_+^n} \mkern-18mu
d\tau_1 \ldots, d\tau_n
\bigg( \sum_{k=1}^n a_k \tau_k\bigg)^s
\prod_{j=1}^n R (\tau_j)\ .
\end{split}
\end{equation}

Next, we recall the notations \eqref{eq5.3}, \eqref{eq6.6},
\eqref{spec-func}
\begin{equation*}
\begin{split}
\lambda (p) & = \int_0^\infty da\, K(a) a^p = \sum_{n=1}^N \alpha_n
\int_{\real_+^n} da_1\ldots da_n A_n (a_1,\ldots, a_n) \bigg(
\sum_{k=1}^n a_k^p\bigg)\ ,
\end{split}
\end{equation*}
and
\begin{equation*}
\begin{split}  \mu (p) & = \frac{
\lambda(p) -1}p\ ,
\end{split}
\end{equation*}
then  Eq.~\eqref{eq10.12} equation can be written in the form
\begin{equation}\label{eq10.15}
(s\mu (1) + \lambda (s) - 1) m_s
= \sum_{n=2}^N \alpha_n I_n (s)\ ,
\end{equation}
where
\begin{equation}\label{eq10.16}
\begin{split}
&I_n (s) = \int_{\real_+^n}\mkern-18mu  da_1\ldots da_n A (a_1,\ldots,a_n)
\int_{\real_+^n}  \mkern-18mu  d\tau_1\ldots, d\tau_n\; g_n^{(s)}
(a_1\tau_1,\ldots, a_n \tau_n) \prod_{j=1}^n R(\tau_j)\\
\noalign{\vskip6pt}
&g_n^{(s)} (y_1,\ldots, y_n)
= \bigg( \sum_{k=1}^n y_k\bigg)^s
- \sum_{k=1}^n y_k^s\ ,\qquad n =1,2,\ldots\ .
\end{split}
\end{equation}

We note that $g_1^{(s)} =0$ for any $s\ge0$ and that $m_0 = m_1=1$
(see Eqs.~\eqref{eq10.8}). Our aim is to study the moments $m_s$
\eqref{eq10.13}, $s>1$, on the basis of Eq.~\eqref{eq10.15}. The
approach is related to the one used in \cite{PT05} for a simplified
version of Eq.~\eqref{eq10.15} with $N=2$. The main results are
formulated below in terms of the spectral function $\mu (p)$ (see
Fig.1) under the assumption that $p_0>1$.

\begin{prop}\label{prop10.1}
$\quad$
\begin{itemize}
\item[{\bf [i]}] If the equation $\mu (s) = \mu (1)$ has the only solution $s=1$,
then $m_s <\infty$ for any $s>0$.
\item[{\bf [ii]}] If this equation has two solutions $s=1$ and $s= s_* >1$,
then $m_s <\infty$ for $s<s_*$ and $m_s =\infty$ for $s>s_*$.
\item[{\bf [iii]}] $m_{s_*} <\infty$ only if $I_n (s_*)=0$ in Eq.~\eqref{eq10.15}
for all $n=2,\ldots,N$.
\end{itemize}
\end{prop}

\begin{proof}
The proof is based on the inequality
\begin{equation}\label{eq10.17}
0 \le \sum_{n=2}^N \alpha_n I_n (s) \le C_N(s) m_1 m_{s-1}
\end{equation}
with $s>1$ and some positive constant $C_N(s)$.
Then we obtain
\begin{equation*}
m_s \le \frac{C_N(s)}{s[\mu (1) - \mu (s)]} \ m_{s-1}\ ,\qquad
m_0 = m_1 = 1\ .
\end{equation*}
In the case [i] $\mu (1) > \mu (s)$ (see Fig.1) for all $s$.
The same is true in the case [ii] for $s< s_*$.
It is clear from Eq.~\eqref{eq10.13} that $m_s>0$ since
$R(\tau)\ge0$ and $m_1 =1$.
This means that the inequality \eqref{eq10.17} cannot be
satisfied for $s>s_*$, therefore moments of orders $s>s_*$
cannot exist.
The statement [iii] follows directly from Eq.~\eqref{eq10.15}.
\end{proof}

Hence, it remains to prove the inequality \eqref{eq10.17}.
The proof is based on the following elementary inequality.

\begin{lem}\label{lem10.2}
In the notation of Eqs.~\eqref{eq10.16},
\begin{equation}\label{eq10.18}
0 \le g_n^{(s)} (y_1,\ldots,y_n)
\le 2^{s-1} s\bigg\{ \psi (y_1,y_2) +\sum_{k=2}^{n-1}
\gamma_k \psi (y_{k+1}, Y_k)\bigg\}\ ,
\qquad s>1\ ,
\end{equation}
where the second term (sum) is absent for $n=2$ and
\begin{equation*}
\begin{split}
& \psi(y_1,y_2) = y_1^{s-1} y_2 + y_2^{s-1} y_1\ ,\qquad
Y_k = \max (y_1,\ldots, y_k)\ ,\\
\noalign{\vskip6pt}
&\gamma_k = \max (k,k^{s-1})\ ,\qquad
k= 1,\ldots,n\ ;\ n= 2,3,\ldots
\end{split}
\end{equation*}
\end{lem}

\begin{proof}
If $n=2$, we assume without loss of generality that $y_1 \le y_2$
and reduce the problem (upper estimate) to the inequality
\begin{equation*}
\Delta (x) = (1+s)^s -1 -x^s - 2^{s-1} s(x+x^{s-1})\le 0\ ,\qquad
x = \frac{y_1}{y_2} \le 1\ .
\end{equation*}
Its proof is obvious since, $\Delta (0)=0$, $\Delta'(x) \le 0$.
The lower estimate in Eqs.~\eqref{eq10.18} is similarly reduced
for $n=2$ to the  inequality
\begin{equation*}
g(\theta) = 1-\theta^s - (1-\theta)^s \ge 0\ ,\qquad
\theta = \frac{x}{x+y} \le 1\ .
\end{equation*}
Then, its proof follows from the fact that $g(0) = g(1) =0$,
$g''(\theta ) \le 0$.

We proceed by induction.
It is easy to see that
\begin{equation*}
g_{n+1}^{(s)} (y_1,\ldots,y_{n+1})
= g_n^{(s)} (y_1,\ldots,y_n) + g_2^{(s)}
\bigg( y_{n+1} , \sum_{k=1}^n y_k\bigg)\ ,\qquad n=3,\ldots\ .
\end{equation*}
Then the lower estimate \eqref{eq10.18} becomes obvious
for any $n\ge2$.
By applying the upper estimate \eqref{eq10.18} for $g_2(s)$
we obtain
\begin{equation*}
g_{n+1}^{(s)} (y_1,\ldots,y_{n+1})
\le g_n^{(s)} + 2^s s\psi \bigg( y_{n+1},\sum_{k=1}^n y_k\bigg)
\end{equation*}
and note that $\psi (x,y)$ is an increasing function of $y$.
Obviously
\begin{equation*}
\sum_{k=1}^n y_k \le n Y_n\ ,\qquad
Y_n = \max (y_1,\ldots, y_n)\ ,
\end{equation*}
and therefore
\begin{equation*}
g_{n+1}^{(s)} (y_1,\ldots, y_{n+1}) \le
g_n^{(s)} (y_1,\ldots, y_n) + 2^s s \gamma_n \psi (y_{n+1},Y_n)\ .
\end{equation*}
This is precisely what is needed to get the upper estimate
\eqref{eq10.18} by induction.
This completes the proof of the lemma.\qed

In order to complete the proof of  the inequality~\eqref{eq10.17} we
substitute the estimates \eqref{eq10.18} into the right hand side
of Eq.~\eqref{eq10.15}.
Then the lower estimate \eqref{eq10.17} becomes  obvious.
The upper estimate \eqref{eq10.17} also becomes obvious
if we note that
\begin{equation*}
\max (a_1\tau_1,\ldots,a_k\tau_k)
\le \bar a_n \max (\tau_1,\ldots,\tau_k)\ ,\quad
\bar a_n = \max (a_1,\ldots,a_n)\ ,\ k=1,\ldots,n\ ;
\end{equation*}
and
\begin{equation*}
\begin{split}
&\int_{\real_+^k} d\tau_1\ldots d\tau_k
\bigg( \prod_{j=1}^k R(\tau_j)\bigg)
[\max  (\tau_1,\ldots,\tau_k)]^s a=\\
\noalign{\vskip6pt}
&\qquad = n!\int_{0\le \tau_1\le\cdots\le \tau_k<\infty }
\mkern-64mu
d\tau_1\ldots d\tau_k \bigg( \prod_{j=1}^k R(\tau_j)\bigg)
\tau_k^s \le n! m_s\ ,\qquad s>0\ .
\end{split}
\end{equation*}
Then it is trivial to show that the upper estimate \eqref{eq10.17} holds.
This completes the proof of Proposition~\ref{prop10.1}
\end{proof}

Now we can draw some conclusions concerning the moments of
the distribution functions \eqref{eq10.11}.
We denote
\begin{equation*}
\begin{split}
m_s (\Phi_p) & = \int_0^\infty dv\, \Phi_p (v) v^s\ ,\qquad
m_s (R_p) = \int_0^\infty d\tau\, R_p (\tau) \tau^s\ ,\\
\noalign{\vskip6pt}
m_{2s} (F_p) & = \int_{\real^d} dv\, F_p (|v|) |v|^{2s}\ ,\qquad
s>0\ ,\ 0< p\le 1\ ,
\end{split}
\end{equation*}
and use similar notations for $N_p (v)$ and $M_p(|v|)$ in
Eqs.~\eqref{eq10.11}.
Then, by formal integration of Eqs.~\eqref{eq10.11}, we obtain
\begin{equation*}
\begin{split}
m_s (\Phi_p) & = m_s (N_p) m_{s/p} (R_p)\\
\noalign{\vskip6pt}
m_{2s} (F_p) & = m_{2s} (M_p) m_{s/p} (R_p)\ ,
\end{split}
\end{equation*}
where $M_p$ and $N_p$ are given in Eqs.~\eqref{eq10.10}.

First we consider the case $0<p<1$.
It follows from general properties of infinitely divisible
distributions that the moments $m_s (N_p)$ and $m_{2s} (M_p)$,
$0<p<1$, are finite if and only if $s<p$  (see \cite{Fe}).
On the other hand, $m_0 (R_p) = m_1 (R_p) =1$, therefore
$m_s (R_p)$ is finite for any $0\le s\le 1$.
Hence, in this case $m_s (\Phi_p)$ and $m_{2s}(F_p)$ are
finite only for $s<p$.

The remaining case $p=1$ is less trivial since {\em all\/}
moments of functions
\begin{gather*}
M_1 (|v|) = (4\pi)^{-d/2} \exp \Big[ - \frac{|v|^2}4\Big] \ ,\qquad
v\in \real^d\ ;\\
\noalign{\vskip6pt}
N_1 (v) = \delta (v-1)\ ,\qquad v\in \real_+\ ,
\end{gather*}
are finite.
Therefore everything depends on moments of $R_1$ in
Eqs.~\eqref{eq10.12} with $p=1$.
It remains to apply Proposition~\ref{prop10.1}.
Hence, the following statement is proved for the moments of
the distribution functions \eqref{eq10.4}, \eqref{eq10.6}.

\bigskip

\begin{prop}\label{prop10.3}
$\quad$
\begin{itemize}
\item[{\bf [i]}] If $0<p<1$, then $m_{2s} (F_p)$ and $m_s (\Phi_p)$
are finite if and only if $0<s<p$.
\item[{\bf [ii]}] If $p=1$, then Proposition~\ref{prop10.1} holds for
$m_s = m_{2s} (F_1)$ and for $m_s = m_s (\Phi_1)$.
\end{itemize}
\end{prop}

\begin{remark}
Proposition~\ref{prop10.3} can be interpreted in other words:
the distribution functions $F_p(|v|)$ and
$\Phi_p(v)$, $0<p\le 1$, can have finite moments of all orders
in the only case when two conditions are satisfied
\begin{itemize}
\item[{\bf{\rm (1)}}] $p=1$,  and
\item[{\bf{\rm (2)}}] the equation $\mu (s)  = \mu (1)$ (see Fig.1)
has the unique solution $s=1$.
\end{itemize}
In all other cases, the maximal order $s$ of finite moments
$m_{2s} (F_p)$ and $m_s (\Phi_p)$ is bounded.
\end{remark}

This fact  means that the distribution functions
$F_p$ and $\Phi_p$ have power-like tails.

\vskip.5in
%%%%%%% %%%%%%%%%%%%%%%%%%%%%%%%%%%%%%%%%%%%%%%%%%%%
%%%%%%%%%%%%%%%    section 11  %%%%%%%%%%%%%%%%%%%%
%%%%%%%%%%%%%%%%%%%%%%%%%%%%%%%%%%%%%%%%%%%%%%%%%%%%

 \section{Applications to the Boltzmann equation}

We recall the three specific Maxwell models {\bf (A)}, {\bf (B)}, {\bf (C)}
 of the  Boltzmann equation from section 2. Our goal in this section
is to study isotropic solutions  $f( |v |, t), v\in\real^d$,
 of Eqs.~\eqref{eq2.2}, \eqref{eq2.4}, and
 \eqref{eq2.5} respectively.
 All three cases are considered below from a  unified point of view.
 First we perform the Fourier transform and denote
\begin{equation}\label{eq11.1}
\begin{split} u(x,t) = \F [f(|v|,t)] = \int_{\real^{d}} dv\, f(|v|,t)
e^{-ik\cdot v}, \quad x=|k|^2, \ \ u(0,t)=1.
\end{split}
\end{equation}
It was already said at the beginning of section~4 that $u(x,t)$
satisfies (in all three cases) Eq.~\eqref{eq4.2}, where $N=2$
 and all notations are given in Eqs.\eqref{eq4.5},
 \eqref{eq3.2}, \eqref{eq3.6}-\eqref{eq3.9}.
 %(3.2), (3.6) - (3.9)
 Hence, all results of our general theory are applicable to these specific
 models. In all three cases  {\bf (A)}, {\bf (B)}, {\bf (C)}
 we assume that the initial distribution function
\begin{equation}\label{eq11.2}
 f(|v|,0)= f_0(|v|) \geq 0\, ,\qquad  \int_{\real^{d}} dv\, f_0(|v|)=1\, ,
\end{equation}
and the corresponding characteristic function
\begin{equation}\label{eq11.3}
u(0,t)= u_0(x) = \F[f_0(|v|)]\, ,\quad  x=|k|^2 \, ,
\end{equation}
are given. Moreover, let $u_0(x)$ be such that
\begin{equation}\label{eq11.4}
 u_0(x) = 1-\alpha x^p + O(x^{p+\ep}),\qquad x\to 0, \qquad  0<p\leq 1,
\end{equation}
with some $\alpha>0$ and $\ep>0$.
We distinguish below the initial data with finite energy (second moment)
\begin{equation}\label{eq11.5}
 E_0 = \int_{\real^{d}} dv\,|v|^2 f_0(|v|) <\infty
\end{equation}
implies $ p=1$ in Eqs.~\eqref{eq11.4}, as follows from Eq.~\eqref{eq10.9},
 and the in-data with infinite energy $E_0=\infty$.

Then $p<1$
 in Eqs.~\eqref{eq11.4} and therefore
\begin{equation}\label{eq11.6}
m_q^{(0)}= \int_{\real^{d}} dv\, f_0(|v|) |v|^{2q} <\infty
\end{equation}
only for $q\leq p<1$ (see \cite{Fe, Lu}).
The case $p>1$ in Eqs.~\eqref{eq11.4}  is impossible for $f_0(|v|)\geq 0$.
 Note that the coefficient $\alpha>0$ in Eqs.~\eqref{eq11.4}
can always be changed to $\alpha=1$ by the scaling transformation
$\tilde x =\alpha^{1/p} x$. Then, without loss of generality,
 we set $\alpha=1$ in \eqref{eq11.4}.

Since it is known that the operator $\Gamma(u)$ in all three cases belongs
to the class \eqref{eq4.3}, we can apply Theorem~\ref{thm8.3}
 and state that the self-similar solutions of Eq.~\eqref{eq4.2}
are given by
\begin{equation}\label{eq11.7}
u_s(x,t) = \Psi(x\, e^{\mu(p) t}), \ \ \ \Psi(x)=w(x^p)\, ,
\end{equation}
where $w(x)$ is given in theorem~\ref{thm8.3} and $0<p<p_0$
(the spectral function $\mu(p)$, defined in \eqref{spec-func},
and its critical point $p_0$ depends on the specific model.)

According to Sections 9-10, we just need to find the spectral function
 $\mu(p)$. In order to do this we first define the linearized
 operator $L=\Gamma'(1)$ for $\Gamma(u)$ given in
Eqs.~\eqref{eq4.3}, \eqref{eq4.5}.
One should be careful at this point since
$A_2(a_1,a_2)$ in Eqs.~\eqref{eq4.5}  is not symmetric and therefore
Eqs.~\eqref{eq4.13}
cannot be used. A straight-forward computation leads to
 \begin{equation}\label{eq11.8}
 \begin{split}
 L\, u(x) & = \int_0^1 ds\, G(s) (\, u(a(s)x) + u(b(s)x)\, )
 +
 \int_0^1 ds H(s)  u(c(s)x) \ ,
 \end{split}
 \end{equation}
in the notation \eqref{eq3.6} -  \eqref{eq3.9}. Then,
 the eigenvalue $\lambda(p)$ is given by
 \begin{equation}\label{eq11.9}
 \begin{split}
L\, x^p &  =\lambda(p)\, x^p\ \ \ \text{which implies}\\
 \lambda(p) & = \int_0^1 ds\, G(s) \left\{  (a(s))^p + (b(s))^p
 \right\}+
 \int_0^1 ds H(s)  (c(s))^p \ ,
 \end{split}
 \end{equation}
and the spectral function  \eqref{spec-func} reads
 \begin{equation}\label{eq11.10}
 \begin{split}
\mu(p)=\frac{\lambda(p) -1 }{p}\, .
 \end{split}
 \end{equation}
Note that the normalization \eqref{eq4.1} is assumed.

At this point we consider the three models {\bf (A), (B), (C) }
separately and apply Eqs.~\eqref{eq11.9} and
 \eqref{eq11.10} to each case.

\noindent {\bf (A) } Elastic Boltzmann Equation~\eqref{eq2.2} in $\real^d, d\geq 2$.
By using Eqs.~\eqref{eq3.6}, \eqref{eq3.7}, and \eqref{eq4.1}
we obtain
 \begin{equation}\label{eq11.11}
 \begin{split}
 \lambda(p) = \int_0^1 ds\, G(s) ( s^p + (1-s)^p ) \, ,
\qquad G(s) =A_d\, g(1-2s) [s(1-s)]^{\frac{d-3}2}\, ,
 \end{split}
 \end{equation}
where the normalization constant $A_d$ is such that Eq.~\eqref{eq4.1}
is satisfied with $H=0$. Then
 \begin{equation}\label{eq11.12}
 \begin{split}
 \mu(p) = \frac1p\int_0^1 ds\, G(s) ( s^p + (1-s)^p - 1) \, , p>0.
 \end{split}
 \end{equation}
It is easy to verify that
 \begin{equation}\label{eq11.13}
 \begin{split}
 &\mu(p) >0 \ \ \text{if}\ p<1\,; \qquad  \qquad\mu(p) <0
\ \ \text{if}\ p>1\,;\\
&\mu(1)=0\, ,\ \ \ \ \ \ \ \qquad \qquad
\mu(2)=\mu(3)= -\int_0^1 ds\, G(s) \, s (1-s) \, ,\\
 &p\,\mu(p) \to 1  \ \ \text{if}\ p\to 0\, ,\qquad \qquad
\mu(p) \to 0  \ \ \text{if}\ p\to\infty\, .
\end{split}
 \end{equation}
 Hence, $\mu(p)$ in this case is similar to the function shown on Fig.1
{\bf b)} with $2<p_0<3$ and such that $\mu(1)=0$. The self-similar
 asymptotics holds, therefore for all $0<p<1$.

\noindent {\bf (B) }{\bf Elastic Boltzmann Equation in the presence of a
thermostat~\eqref{eq2.4} in $\real^d, d\geq2$.}
We consider just the case of a cold thermostat with $T=0$ in
Eq.~\eqref{eq3.2}, since the general case $T>0$ can be considered after
 that with the help of \eqref{eq2.11}.
Again, by using Eqs.~\eqref{eq3.6}, \eqref{eq3.7}, and \eqref{eq4.1}
we obtain
 \begin{equation}\label{eq11.14}
 \begin{split}
 \lambda(p) &= \int_0^1 ds\, G(s) ( s^p + (1-s)^p ) +
\theta\, \int_0^1 ds\, G(s) \big( 1- \frac{4m}{(1+m)^2} \big)^p ,\\
 G(s) & = \frac{1}{1+\theta}\, A_d\, g(1-2s) [s(1-s)]^{\frac{d-3}2}\, ,
 \end{split}
 \end{equation}
with the same  constant $A_d$ as in Eq.~\eqref{eq11.11}. Then
 \begin{equation}\label{eq11.15}
 \begin{split}
 \mu(p) &= \frac{1}{p}\int_0^1 ds\, G(s)
( \, s^p + (1-s)^p - \theta(1-\beta s)^p
-(1+\theta) \, )\, , \\
 \beta &= \frac{4m}{(1+m)^2}\, , \ \  p>0,
 \end{split}
 \end{equation}
and therefore
 \begin{equation}\label{eq11.16}
 \begin{split}
\  &\mu(1)=  -\theta\, \beta\,\int_0^1 ds\, G(s) \, s  \, ,\\
\  &p\,\mu(p) \to 1  \ \ \text{if}\ p\to 0\, \qquad \qquad
\mu(p) \to 0  \ \ \text{if}\ p\to\infty\, ,
\end{split}
 \end{equation}
which again verifies that  $\mu(p)$
is of the same kind as in the elastic case {\bf A)} and shown on
Fig.1~{(b)}.
A position of the critical point $p_0$ such that $\mu'(p_0)=0$
(see Fig.1~{(b)})  depends on $\theta$. It is important to distinguish
two cases: {\bf 1)} $p_0>1$ and  {\bf 2)} $p_0<1$. In case   {\bf 1)} any
 non-negative initial data \eqref{eq11.2} has the self-similar asymptotics.
In  case  {\bf 2)} such asymptotics holds just for in-data with infinity
 energy satisfying Eqs.~\eqref{eq11.4} with some $p<p_0<1$. A simple
 criterion to separate  the two cases follows directly from   Fig.1~{(b)}:
it is enough to check the sign of $\mu'(1)$. If
 \begin{equation}\label{eq11.17}
\mu'(1) =\lambda'(1) -\lambda(1) +1 <0
 \end{equation}
in the notation of Eqs. \eqref{eq11.14}, then $p_0>1$ and the
self-similar asymptotics holds for any non-negative initial data.

The inequality \eqref{eq11.17} is equivalent to the following condition on
 the positive coupling constant $\theta$
 \begin{equation}\label{eq11.18}
 \begin{split}
0 < \theta<\theta_* = - \frac{\int_0^1 ds\, G(s)\ (\, s \log s +
(1-s)\log(1-s)\, )}
{ \int_0^1 ds\, G(s)\ (\, \beta \, s  + (1-\beta s)\log(1-\beta s)\, ) }\, .
\end{split}
 \end{equation}
The right hand side of this inequality is positive and independent on the
 normalization of $G(s)$, therefore it does not depend on $\theta$
(see Eq.~\eqref{eq11.15}. We note that a new class of exact self-similar
solutions to Eq.~\eqref{eq2.4} with finite energy was recently found in
\cite{BG-06} for $\beta=1, \theta=4/3$ and $G(s)=const.$
A simple calculation of the
integrals in \eqref{eq11.18} shows that $\theta_*=2$ in
that case, therefore the criterion \eqref{eq11.17} is fulfilled
for the exact solutions from \cite{BG-06} and they are asymptotic for a wide
 class of initial data with finite energy.
Similar conclusions can be made in the same way about exact positive
self-similar solutions with infinite energy
 constructed in \cite{BG-06}. Note that the
 inequality \eqref{eq11.18}  shows the non-linear character of the
self-similar asymptotics: it holds  unless the linear
 term in  Eq.~\eqref{eq2.4}
 is `too large'.

\noindent {\bf (C) } Inelastic Boltzmann Equation~\eqref{eq2.5} in $\real^d$.
Then Eqs.~\eqref{eq3.9} and \eqref{eq4.1} lead to
 \begin{equation*}
\label{eq11.19n}
 \begin{split}
 &\qquad\lambda(p) = \int_0^1 ds\, G(s) ( (a\,s)^p + (1-b\,s)^p ) \, ,\\
&\text{where}\\
&\qquad G(s) = C_d\,  (1-s)^{\frac{d-3}2}\, ,\qquad a=\frac{(1+e)^2}4 \, ,\ \
\ \ b=\frac{(1+e)(3-e)}4\, ,
 \end{split}
 \end{equation*}
with such constant $C_d$ that Eq.~\eqref{eq4.1} with $H=0$ is fulfilled.
Hence
 \begin{equation}\label{eq11.19}
 \begin{split}
 \mu(p) = \frac1p\int_0^1 ds\, G(s) (\,  (a\,s)^p + (1-b\,s)^p  - 1) \, ,
\qquad p>0\, ,
 \end{split}
 \end{equation}
and therefore
 \begin{equation*}
 \begin{split}
&\mu(1)=-\frac{1-e^2}4\int_0^1 ds\, G(s) \,s \, , \ \ \ \ \ \ \ \\
 &p\,\mu(p) \to 1  \ \ \text{if}\ p\to 0\, ,\qquad \qquad
\mu(p) \to 0  \ \ \text{if}\ p\to\infty\, .
\end{split}
 \end{equation*}
Thus, the same considerations lead to the shape of $\mu(p)$ shown in
 Fig.1~{(b)}.
The inequality  \eqref{eq11.6} with $\lambda(p)$ given in Eqs.
\eqref{eq11.19} was proved in \cite{BCT-03} (see Eqs. (4.26) of \cite{BCT-03},
where the notation is slightly different from ours). Hence, the
inelastic Boltzmann equation~\eqref{eq2.5}  has self-similar asymptotics for
any restitution coefficient $0<e<1$ and any non-negative initial data.

\medskip

Hence, the spectral function $\mu(p)$ in all three cases above is such
 that $p_0>1$ provided the inequality \eqref{eq11.18} holds for the model
{\bf (B)}.

Therefore, according to our general theory, all `physical' initial
conditions \eqref{eq11.2} satisfying Eqs. \eqref{eq11.4} with any
$0<p\leq 1$ lead to self-similar asymptotics. Hence, the main properties of
the solutions $f(v,t)$ are qualitatively similar for all three models
{\bf (A), (B)} and {\bf (C)}, and  can be described in one unified statement:
Theorem~\ref{thm11.1} below.

Before we formulate such general statement, it is worth to clarify one
 point related to a special value $0<p_1\leq 1$ such that $\mu(p_1)=0$.
The reader can see from Fig.1~{(b)} that the unique root of this
equation exists  for all models {\bf (A), (B),}{\bf (C)}
 since $\mu(1)=0$ in the case  {\bf (A)} (energy conservation) , and
$\mu(1)<0$ in  cases  {\bf (B)} and  {\bf (C)} (energy dissipation).
If $p=p_1$ in Eqs. \eqref{eq11.4}
then the self-similar solution  \eqref{eq11.7} is simply a stationary
solution of Eq. \eqref{eq4.2}. Thus, the time relaxation to the
non-trivial ($u\neq 0,1$) stationary solution is automatically included in
Theorem~\ref{thm11.1} as a particular  case of self-similar asymptotics.

Thus we consider simultaneously Eqs.~\eqref{eq2.2}, \eqref{eq2.4},
\eqref{eq2.5}, with the initial condition \eqref{eq11.2} such that
Eq. \eqref{eq11.4} is satisfied with some $0<p\leq 1$, $\ep >0$
and $\alpha=1$. We also assume that $T=0$ in Eq. \eqref{eq2.4} and the
 coupling parameter $\theta>0$ satisfies the condition \eqref{eq11.18}.

In the following Theorem~\ref{thm11.1},  the solution $f(|v|,t)$ is
understood in each case as a generalized  probability density in
$\real^d$ and the convergence  $f_n\to f$ in the sense of weak
 convergence of probability measures.

\begin{thm}\label{thm11.1} The following two statements hold

\begin{itemize}
\item[{\bf [i]}]  There exists a unique (in the class of probability measures)
solution $f(|v|,t)$ to each of Eqs.~\eqref{eq2.2}, \eqref{eq2.4},
\eqref{eq2.5} satisfying the initial condition  \eqref{eq11.2}. The solution
 $f(|v|,t)$  has self-similar asymptotics in the following sense:

For any given $0<p\leq 1$ in  Eqs.~\eqref{eq11.4}
there exits a unique non-negative self-similar solution
 \begin{equation}\label{eq11.20}
 \begin{split}
 f_s^{(p)}(|v|,t) = e^{-\frac{d}2\mu(p)\,t} F_p(|v|e^{-\frac12{\mu(p)}\, t})\ ,
 \end{split}
 \end{equation}
such that
 \begin{equation}\label{eq11.21}
 \begin{split}
 e^{\frac{d}2\mu(p)\, t} f(|v| e^{-\frac12 \mu(p)\,t},t) \
\to_{t\to\infty}  F_p(|v|)\ ,
 \end{split}
 \end{equation}
where $\mu(p)$ is given in Eqs.~\eqref{eq11.12}, \eqref{eq11.15},
\eqref{eq11.19}, respectively, for each of the three models.
\item[{\bf [ii]}] Except for the special case of the Maxwellian
 \begin{equation}\label{eq11.22}
 \begin{split}
F_1(|v|)=
M(|v|) =(4\pi)^{-d/2} e^{-\frac{|v|^2}4}
 \end{split}
 \end{equation}
for Eq.~\eqref{eq2.2} with $p=1$ in Eq.~\eqref{eq11.4} (note that $\mu(1)=0$ in
this case), the function $ F_p(|v|)$ does not have finite moments of all orders.
If $0<p<1$, then
\begin{equation}\label{eq11.23}
m_q = \int_{\real^d} dv\, F_p(|v|) |v|^{2q} < \infty \qquad\text{only for  }
0<q<p\, .
 \end{equation}
If $p=1$ in the case of Eqs.~\eqref{eq2.4},
\eqref{eq2.5}, then $m_q <\infty$ only for  $0<q<p_*$, where $p_*>1$ is the unique
maximal root of the equation $\mu(p_*)=\mu(1)$, with   $\mu(p)$ given in
Eqs.~\eqref{eq11.4}, \eqref{eq11.19} respectively.
\end{itemize}
\end{thm}
\begin{proof}
It is well known \cite{Fe} that there exists  a one-to-one correspondence
 between probability measures and their characteristics functions
 \eqref{eq11.1}. Moreover, the point-wise convergence of characteristic
 functions is equivalent to the weak convergence of probability measures.
Therefore, the problem for $f(|v|,t)$ can be reduced to study the
problem
 \eqref{eq4.6} for
$u(x,t)$ with the corresponding operator $\Gamma$ as in   \eqref{eq4.3}.
Then,  the statement {\bf [i]} of the theorem follows from the  corresponding
results for the initial value problem  \eqref{eq4.6} (see, in particular,
 Proposition~\ref{prop9.1}). We just need to express these results in terms of
the distribution functions   $f(|v|,t)$
(see Eqs.\eqref{eq10.4}, \eqref{eq10.6}).

Concerning the  statement {\bf [ii]} of the theorem, the exceptional case
 of the Maxwellian \eqref{eq11.23} is clear since $p_*=1$ is the only root of the
 equation $\mu(p_*)=\mu(1)=0 $ with  $\mu(p)$ given in
Eq.~\eqref{eq11.12}. The rest of the statement  {\bf [ii]} follows from
Proposition~\ref{prop10.3}. So the theorem is proved.
\end{proof}

\medskip
In fact we also have some additional information (not included in the
formulation of Theorem~\ref{thm11.1}) about solutions
$f(|v|,t)$. We can express this information in the following way.

\begin{corollary}\label{cor11.2}
Under the same  conditions of Theorem~\ref{thm11.1},
the following two statements hold.
\begin{itemize}
\item[{\bf [i]}] The rate of convergence in Eq. \eqref{eq11.21} is
 characterized in terms of the corresponding characteristic functions
 in Proposition~\ref{prop9.1}.
\item[{\bf [ii]}] The function $ F_p(|v|)$
admits the integral representation \eqref{eq10.11} through infinitely
divisible distributions  \eqref{eq10.10}.
\end{itemize}
\end{corollary}
\begin{proof}
It is enough to note that all results of section 9 and 10 are valid,
 in particular, for Eqs.~\eqref{eq2.2}, \eqref{eq2.4},
\eqref{eq2.5}.
\end{proof}

\begin{remark} To complete our presentation, we mention that
a statement similar to theorem~\ref{thm11.1},
 can be easily derived  from general results of sections 9 and 10 in the case
 of $1-d$ Maxwell models introduced in \cite{To-05}, \cite{PT05}, \cite{BN05}
for applications to economy models (Pareto tails, etc.). The only difference
 in that case is that the `kinetic' equation can be transformed to its canonical form
 \eqref{eq4.2}-\eqref{eq4.3} by the Laplace transform as discussed in
 section 10, and that the corresponding spectral function $\mu(p)$ can have
 any of the four kind of behaviors shown in Fig.1. Therefore, the only remaining
 problem for any such $1\, d$ models is to study them for their  specific
function $\mu(p)$, and then to apply
 Propositions~\ref{prop9.1}, \ref{prop10.1}
 and \ref{prop10.3}.
Thus, the general theory developed in the present paper is
applicable to all
 existing multi-dimensional isotropic Maxwell models and to $1\,d$ models as well.
\end{remark}
\vskip.5in
%%%%%%%%%%%%%%%%%%%%%%%%%%%%%%%%%%%%%%%%%%%%%%%%%%%%%%%%%%%%%%%%%%%%
%%%%%%%%%%%%%%%%%%%%%%%%%%%%%%%%%%%%%%%%%%%%%%%%%%%%%%%%%%%%%%%%%%%5
\section*{Acknowledgements}

The first author was supported by  grant 621-2003-5357 from the
Swedish Research Council (NFR). The research of the  second author
was supported by MIUR of Italy.  The third author
 has been partially supported by NSF under grant
DMS-0507038. Support from the Institute for Computational
Engineering and Sciences at the University of Texas at Austin is
also gratefully acknowledged.

\vskip.5in

%%%%%%%%%%%%%%%

 \end{document}